\documentclass[10pt,twocolumn,oneside,journal]{IEEEtran}

\ifCLASSINFOpdf 
\usepackage[pdftex]{graphicx}

\graphicspath{figures/}
\DeclareGraphicsExtensions{.pdf,.jpeg,.png}
\else
\fi

\usepackage{cite}
\usepackage[cmex10]{amsmath}
\usepackage{amsfonts}
\usepackage{amssymb}
\usepackage{algorithm}
\usepackage{algorithmic}
\usepackage{siunitx}
\usepackage{bm}
\usepackage[caption=false,font=footnotesize]{subfig}

\usepackage{array}
\usepackage{lipsum}
\setlipsumdefault{1-2}
\usepackage{url}
\usepackage{hyperref}    

\usepackage{fixme}
\usepackage{pdfcomment}
\usepackage{lipsum}
\usepackage{todonotes}
 \usepackage{tikz,pgfplots}
 \usepackage{etex}
   \pgfplotsset{compat=newest} 
  \pgfplotsset{plot coordinates/math parser=false}
\fxsetup{final}
\usepackage{psfrag}
\usepackage{multirow}
\usepackage[pages=some,placement=top]{background}
\SetBgContents{ \footnotesize This work has been submitted to the IEEE
 for possible publication. Copyright may be transferred without
 notice, after which this version may no longer be accessible.}
\SetBgScale{1}
\SetBgColor{black}

\begin{document}

% include the body
\BgThispage
% LaTeX source code for the article 
% Model-Based Calibration of Filter Imperfections in
% the Random Demodulator for Compressive Sensing
% Written by Pawel Jerzy Pankiewicz, Aalborg University
% email: pjp@es.aau.dk
\title{Model-Based Calibration of Filter Imperfections in the Random Demodulator for Compressive Sensing}

\author{Pawel~J.~Pankiewicz, Thomas Arildsen,~\IEEEmembership{Member, IEEE}, and~Torben Larsen,~\IEEEmembership{Senior Member, IEEE}

\thanks{The authors are with Aalborg University, Faculty of Engineering and
Science, Department of Electronic Systems, DK-9220 Aalborg, Denmark. The
authors' e-mails are: \{pjp, tha, tl\}@es.aau.dk. This work was financed by The Danish Council for Strategic Research under grant number 09-067056. The authors would like to thank Danish Center for Scientific Computing (DCSC) for funding.
}
}

\maketitle
\begin{abstract}
	\boldmath
	The random demodulator is a recent compressive sensing architecture providing efficient sub-Nyquist sampling of sparse band-limited signals. The compressive sensing paradigm requires an accurate model of the analog front-end to enable correct signal reconstruction in the digital domain. In practice, hardware devices such as filters deviate from their desired design behavior due to component variations. 
	Existing reconstruction algorithms are sensitive to such deviations, which fall into the more general category of measurement matrix perturbations.
  This paper proposes a model-based technique that aims to calibrate  filter model mismatches to facilitate improved signal reconstruction quality. The mismatch is considered to be an additive error in the discretized impulse response. We identify the error by sampling a known calibrating signal, enabling least-squares estimation of the impulse response error. The error estimate and the known system model are used to calibrate the measurement matrix. Numerical analysis demonstrates the effectiveness of the calibration method even for highly deviating low-pass filter responses. The proposed method performance is also compared to a state of the art method based on discrete Fourier transform trigonometric interpolation.
\end{abstract}


\IEEEpeerreviewmaketitle

\section{Introduction}

% ############################################################
% ###########----- SYMBOLS definitions ----###################
\newcommand{\x}{\mathbf{x}}
\newcommand{\y}{\mathbf{y}}
\newcommand{{\bPhi}}{\bm{\Phi}}
\newcommand{\bPsi}{\bm{\Psi}}
\newcommand{\balpha}{\bm{\alpha}}
\newcommand{\bH}{\mathbf{H}}
\newcommand{\bP}{\mathbf{P}}
\newcommand{\bD}{\mathbf{D}}
\newcommand{\Mq}{$M_{\text{q}}$}
\newcommand{\Rl}{R_{\text{l}}}
\newcommand{\Rs}{R_{\text{s}}}

\renewcommand{\algorithmicrequire}{\textbf{Input:}}
\renewcommand{\algorithmicreturn}{\textbf{Output:}}

% TIKZ size setting
    \newlength\figureheight 
    \newlength\figurewidth
% ####################----###################----############

% background
\IEEEPARstart{T}{he} compressive sensing (CS) paradigm~\cite{Candes2006b, Donoho:2006vb, Candes:2008uc} has inspired researchers to apply the theory in practical analog signal acquisition~\cite{Kirolos:2006p3806, Laska:2007p3817, Mishali:2010p3825, Mishali:2010p4876, Ragheb:2008vb, Yang:2009vb, becker2011 }. An analog-to-digital converter (ADC) utilizing the CS framework can sample sparse or compressible signals at significantly lower frequencies than the Shannon-Nyquist theory for general and potentially dense signals dictates~\cite{Unser:2000vo, Shannon:1949:CPN}. 
The Shannon-Nyquist condition is a sufficient sampling criterion when no prior information on the signal composition is available. Following the principles of CS, the under-sampled signal can be reconstructed if it is sparse or compressible. Signal sparsity is modeled by expressing the signal as the linear combination of a few elements from a particular dictionary~\cite{Candes2006b}. The trade-off in CS is a more complex signal recovery as it requires non-linear reconstruction algorithms~\cite{tropp2010c}.   

% background - RD description
The random demodulator (RD) sampling architecture has been widely explored since the introduction of the compressive sensing theory~\cite{Kirolos:2006p3806, Laska:2007p3817, Tropp:2010p3813, Ragheb:2008vb}. The architecture is dedicated to the sampling of frequency-, time-frequency- or time-sparse signals~\cite{Anonymous:ctmnJhsb, becker2011, Kirolos:2006p3806} which makes it more flexible than other analog CS architectures such as~\cite{Mishali:2010p3825, Mishali:2010p4876, Yang:2009vb}. The acquisition process leads to fewer samples than the traditional Shannon-Nyquist method.

The RD architecture illustrated in Fig.~\ref{fig:rd-architecture}, can be implemented by standard off-the-shelf components~\cite{Tropp:2010p3813, Kirolos:2006p3806}. The RD architecture aim is to compress an analog input signal into a smaller bandwidth, which can be further sub-sampled, encoding the signal information on smaller set of samples. The core idea behind the compression in the RD architecture is to modulate the input signal by a fast-varying chipping sequence and to low-pass filter it. The sub-sampling operation is realized by a low-rate sampling ADC. These functional procedures are modeled by the so-called measurement matrix in CS signal reconstruction algorithms~\cite{Tropp:2010p3813, becker2011}. The reconstruction relies on the accuracy of the measurement matrix\cite{eldar2012compressed}.

% Problem statement
In reality, due to factors such as supply voltage, manufacturing process, temperature variations etc., the analog components do not behave ideally and hence the actual front-end differs from its ideal model. 
Due to the relatively low clock rate of the RD some imperfections such as clock jitter and nonlinear distortion can be neglected~\cite{Tropp:2010p3813}. However, stationary imperfections such as component impairments cannot be neglected~\cite{pankiewicz11}. Previous studies show that generic CS reconstruction algorithms are sensitive to mismatches between the ideal and the actual analog front-end, represented by the measurement matrix~\cite{pankiewicz11, Gribonval12}. The need for measurement matrix calibration has therefore been emphasized in~\cite{Tropp:2010p3813, Ragheb:2008vb}.

% State-of-the art
An obvious solution, although impractical, is to measure the actual impulse response of each device and revise the model (measurement matrix) accordingly~\cite{Ragheb:2008vb}.
Existing literature also investigates the question of how much error the mismatch in the measurement matrix contributes to the reconstruction quality~\cite{Herman2010, Herman:2010td, Wang:2011kc }. This is, however, an analysis of the problem -- not an attempt to mitigate it.
Several proposals of a more robust reconstruction  have also been made \cite{Rosenbaum2010, Zhu2011, Liu:2010tc, Han:2011eh}. The algorithms consider an additive  error in the measurement matrix or dictionary. This enables a more robust signal estimate, assuming only statistical knowledge of the error. 

In~\cite{becker2011}, the author discusses calibration of an analog CS architecture based on the RD. The methodology considered building the system's measurement matrix via the Fourier domain by sampling specially dedicated signal sequences. The technique is known as discrete Fourier transform trigonometric interpolation (DFTTI)~\cite{becker2011}. The method is accurate and does not require an initial front-end model, although   depending on the systems' parameters, the DFTTI might be time-consuming. The operation requires calibrating samples of the same order as the CS measurement matrix problem size $M \times N$ ($M<N$),  where $M$ denotes compressed samples and $N$ the amount of Nyquist samples of the input signal.  
Also, a blind sparse calibration of an initially modeled measurement matrix has been proposed \cite{Gribonval12}. The method calibrates the measurement matrix through $M$ samples from $U$ unknown (but sparse) training sequences. The procedure requires $U\times M$ calibrating measurements, where $U<<M$. 

% the proposal
This article proposes a supervised model-based calibration method that minimizes the discrepancy between the initially modeled measurement matrix and the actual front-end. The method exploits the nature of the error associated with the measurement matrix through sampling of an a-priori known signal to identify the errors through linear estimation. The error estimate is further used to calibrate the initially modeled measurement matrix. The method can be seen as a trade-off between the sample-expensive DFTTI supervised method and sample-efficient unsupervised sparse calibration. The successful model-based calibration requires only $S$ supervised measurements, where $S\leq M$. In this paper we focus on the practical aspects of the RD architecture, testing the calibration on modeled component impairments. Performed signal reconstruction benchmarks with the DFTTI method~\cite{becker2011} show significant time advantages in favor of our proposed method.

The rest of the paper is structured as follows: Section~\ref{sec:theframework} presents the RD  and CS frameworks. Section~\ref{sub:measurement_matrix} describes the measurement matrix structure and the modeling error. Section~\ref{sec:calibration_methodology} presents the proposed  calibration principle. Section~\ref{sec:simulation_framework} describes simulation framework, the case study of a passive filter with imperfect components used in the random demodulator, and  calibration benchmark results.
Finally, section~\ref{sec:conclusion} presents the conclusion.

\section{Background} % (fold)
\label{sec:theframework}
The RD obtains measurements $\y$ according to the CS principle~\cite{Kirolos:2006p3806,  Tropp:2010p3813}: 
\begin{equation}
	\label{eq:cs_primer}
	\y = \bPhi\x, 
\end{equation} 

\noindent where $\bPhi \in \mathbb{R}^{M \times N}$, $M < N$ is the measurement matrix that represents the analog front-end of the random demodulator, $\x \in \mathbb{R}^{N\times 1}$ is the original signal, and $\y \in \mathbb{R}^{M\times 1}$ denotes compressed measurements acquired for time $t\in [0, T)$. $T$ denotes the observation time length. The sampling rate $f_{\text{s}} = M/T$ needed for successful signal recovery is dictated by a lower bound of $M \ge  CK\log_{10}{(\frac{2\mathcal{B}}{K} + 1)}$, rather than $2\mathcal{B}$, where $\mathcal{B}$ is the bandwidth of a signal, $K$ is the signal sparsity, $C$ is a positive constant acquired empirically~\cite{Tropp:2010p3813, eldar2012compressed,Kirolos:2006p3806}.  
A sparse representation is one of the necessary requirements to utilize CS~\cite{Candes2006b, Donoho:2006vb}. A model of a sparse signal can be represented as:
\begin{equation}
	\label{eq:sparsity}
	\x = \bm\Psi \bm{\alpha}, 
\end{equation}   
where $\boldsymbol\Psi$ is an ${N \times N}$ dictionary matrix, and $\balpha$ of size ${N\times 1}$ is the underlying sparse vector, i.e., $\balpha$ contains $K \ll N$ non-zero coefficients. Alternatively, $\balpha$ may be compressible instead. This more relaxed requirement is met when the entries of $\balpha$ decay rapidly to zero when sorted by magnitude.

\begin{figure}[htb]
	\centering
		\includegraphics[width=\columnwidth]{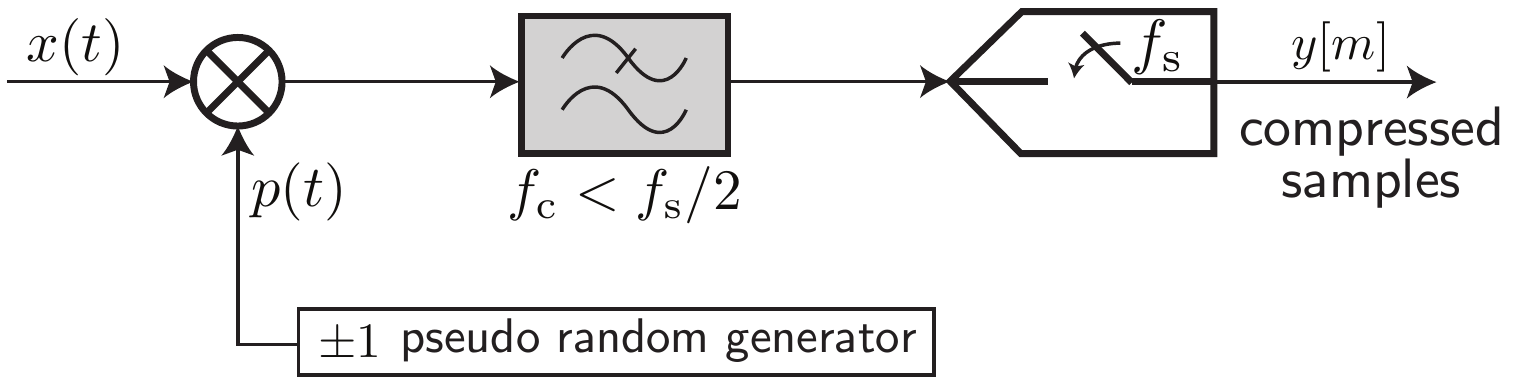}
	\caption{Single stage random demodulator excluding quantization of the compressed measurements\cite{Kirolos:2006p3806}.}
	\label{fig:rd-architecture}
\end{figure}

The RD architecture is illustrated in Fig.~\ref{fig:rd-architecture}. First the analog signal $x(t)$ is spread in frequency by the multiplier and $p(t)$, the signal is low-pass filtered and subsequently uniformly sampled at frequency $f_{\rm s}$.

\begin{figure}[htbp]
	\centering
		\includegraphics[width=\columnwidth]{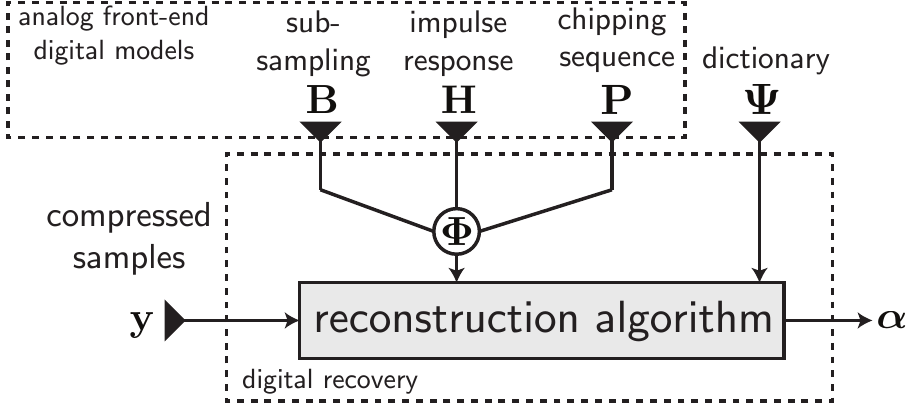}
	\caption{Conceptual illustration of the CS signal reconstruction for the RD technique. }
	\label{fig:rd-reconstruction-stage}
\end{figure}

The compressed measurements $y[0],\ldots, y[M-1]$ are then used to reconstruct the sub-sampled signal by a suitable algorithm, \cite{Candes2006b, Tropp:2010p3813, Laska:2007p3817, Davies:2008dk, tropp2007,
Dai2009b}. The principle is to utilize the compressed measurements $\y$
together with a sparsifying dictionary $\bPsi$ and measurement matrix $\bPhi$ to recover the sampled signal $\x$ as illustrated in~Fig.~\ref{fig:rd-reconstruction-stage}. 

\subsection{Reconstruction stage } % (fold)
\label{sub:reconstruction_algorithms_}
Initially, in order to recover a sampled signal $\x$ from compressed measurements $\y$, we would use the assumption of sparsity~\cite{Candes2006b}.
The problem in a computationally tractable form can be posed as a convex problem, where a sparse vector is recovered as: $\text{argmin}~\|\balpha\|_1~\text{s.t.}~\y=\bPhi\bPsi\balpha$. 
\newcommand{\reals}{\mathop{\mathbb{R}}}
\newcommand{\argmin}{\mathop{\rm argmin}}
\newcommand{\argmax}{\mathop{\rm argmax}}

This approach is called Basis Pursuit \cite{Chen2001} and it belongs to the family
of convex optimization methods used to recover signals within the
CS framework~\cite{Donoho:2006vb}.

More practical reconstruction methods can be constructed under the assumption of noise added to the compressed samples $\y$  as a consequence of the sampling process, e.g., quantization in the ADC. This approach is known as Basis Pursuit De-Noising (BPDN)~\cite{Chen2001}: 
\begin{equation}
 \begin{array}{ll}\label{eq:BPDN}
 \displaystyle \mathop{\mbox{minimize}}_{\balpha \in \mathbb{C}^{N\times 1}} & \|\balpha\|_{1}\\
 \mbox{subject to} & \|\y - \bPhi\bPsi\balpha\|_{2} \leq \zeta, \\
 \end{array}
\end{equation}
where $\zeta$ controls the fidelity term.  

The $\mathcal{\ell}_{1}$-minimization techniques present strong recovery guarantees but suffer from high implementation complexity\cite{tropp2007}.
Apart from convex optimization approaches, there
is a group of methods called greedy algorithms where the unknown support of the signal is calculated iteratively. The Orthogonal Matching Pursuit (OMP)~\cite{tropp2007} and the Subspace Pursuit (SP)~\cite{Dai2009b} are some of the most popular methods in this group. 
% section reconstruction_algorithms_ (end)

\subsection{Functionality of the acquisition stage } % (fold)
\label{sub:system_functionality}
% subsection system_functionality (end)
The RD architecture is dedicated to handling band-limited signals and assumes that the analog signal $\mathit{x(t)}$ is composed of a discrete, finite number of weighted continuous dictionary components as\cite{Kirolos:2006p3806, Laska:2007p3817}:
\begin{equation}
	x\left( t \right) = \sum_{n= 0}^{N-1} \alpha _{n}\psi
   _{n}\left( t \right),~t\in[0,T),
   \label{eq:signal_model}
\end{equation}
where, e.g., $\alpha_{0}, \ldots, \alpha_{N-1} \in \mathbb{C}$ for frequency-sparse signals could represent Fourier series coefficients $\psi_{n}\left( t \right) = \exp \left[-j 2\pi n t\right]$~\cite{Tropp:2010p3813}.

The RD signal acquisition starts with a spread spectrum operation. The operation is carried out by multiplying the input signal by the chipping sequence, produced by a random number generator: 
\begin{equation}
  \label{eq:mixing}
  d(t) = x(t)\,p(t),
\end{equation}
where $p(t)$ is the chipping sequence. The zero-mean $\pm 1$ chipping sequence has to be alternating at the frequency $f_{\rm chip} > 2\mathcal{B}$ of the input signal~\cite{Tropp:2010p3813, becker2011}. According to \cite{becker2011}, it is desirable that $f_{\rm chip}$ is as close as possible to the lower bound to keep most of the power in-band.

The filtering operation can be represented as a convolution of the mixed signal with the impulse response of the filter $h$\cite{Laska:2007p3817}:
\todo[size=\footnotesize,disable]{make cdots consistent}
\begin{equation}
\label{eq:lpf_ex}
  \\ x_{\rm lpf}(t) = \int_{-\infty}^{+\infty} d(\tau)\, h(t- \tau)\,\mathrm{d}\tau.
 \end{equation}

Lastly, the filtered signal $x_{\rm{lpf}}$ is uniformly sampled at the rate $f_{\rm s}$ and  yields compressed measurements $\y \in \mathbb{R}^{M\times 1}$.

The system described by~(\ref{eq:mixing})--(\ref{eq:lpf_ex}) and the sampling are linear operations. Considering the signal model in (\ref{eq:signal_model}), the discrete compressed measurement vector can be characterized as a linear transformation of the discrete coefficient vector $\balpha$. Further expanding~(\ref{eq:lpf_ex}), as shown in~\cite{Kirolos:2006p3806}, results in the following model for the compressed measurements discrete vector:
\begin{equation}
  \label{eq:y-math}
  y[m] = \sum_{n=0}^{N-1} \alpha_{n} \int_{-\infty}^{+\infty}  \psi_{n}(\tau)\, p(\tau)\, h(m f_{\text{s}}^{-1}- \tau)\,\mathrm{d}\tau.
\end{equation}
 
The model of the analog front-end in the reconstruction stage is represented in a digital form and \eqref{eq:y-math} is therefore discretized to the following form\footnote{Assuming that $x(t)$ and $p(t)$ are equal to zero for $t<0$, and impulse response is discretized to $N$ samples.}:
 \begin{equation}
  \label{eq:y-mathd_alpha}
  y[m] \simeq \sum_{v=0}^{N-1} \sum_{n=0}^{N-1} \alpha[v]\: \psi[v,n] \: p[n]\: h[mR- n],
\end{equation}
and by utilizing the sparse model in~\eqref{eq:sparsity}:  
 \begin{equation}
  \label{eq:y-mathd_x}
  \y[m] \simeq \sum_{n=0}^{N-1} x[n] \: p[n] \: h[mR- n],
\end{equation}
 \noindent where $R = {f_{2\mathcal{B}}}/{f_{\text{s}}} = {N}/{M} \in \mathbb{N}^1$ is a positive integer that defines the sub-sampling ratio in discretized form~\cite{eldar2012compressed}.
 The operations on the right hand side of~\eqref{eq:y-mathd_alpha} are expressed using a linear transformation $\y = \bPhi\bPsi\balpha$.  The dictionary and filter matrices entail both time and frequency discretization of the dictionary and time discretization of the filter.  As described in the introduction section, $\bPhi$ is the measurement matrix mapping $\x$ to the compressed set of measurements $\y$, and $\bPsi$ is the sparsity basis with assumption of integer tone separation equal to $1$, in the case of frequency sparse signals\cite{Tropp:2010p3813, eldar2012compressed}.

\section{Measurement matrix structure} % (fold)
\label{sub:measurement_matrix}
The measurement matrix represents a model of the
operations undergone by the signal during acquisition~\cite{Anonymous:ctmnJhsb}. 
From~(\ref{eq:y-math}) and (\ref{eq:y-mathd_x}) we can isolate expressions for modulation, filtering and sampling: 
\begin{equation}
	\label{eq:psi_extended}
	\boldsymbol\Phi = \mathbf{BHP},
\end{equation}
where the matrix $\mathbf{\Phi}$ is considered the product of  three matrix factors representing the uniform sub-sampling $\mathbf{B} \in \mathbb{Z}^{M \times N}$, impulse response of the filter $\mathbf{H} \in \mathbb{R}^{N \times N}$ and chipping sequence $\mathbf{P} \in \mathbb{Z}^{N \times N}$.

The chipping sequence matrix is defined as follows:

\begin{equation}
\label{eq:chip_seq_d}
    \mathbf{P} = {\rm diag}\left(p[0],p[1], \ldots ,p[N-1] \right) \in \{\pm 1\}^{N \times N}, 
\end{equation}
 The spread spectrum operation in~(\ref{eq:mixing}), in the discrete form, is interpreted as a product of $\x$ and $\mathbf{P}$, that yields $N$  demodulated samples:
\begin{equation}
	\label{eq:demodulation-digital}
	d[n] = x[n]\ p[n],~n \in \{0, \ldots, N-1\}.
\end{equation}

The matrix representing an approximation of the infinite impulse response of the filter or more generally linear time invariant (LTI) system has the form of a banded $N\text{-by-}N$ Toeplitz matrix:
\small{
\begin{equation}
\label{eq:H-matrix}
	\mathbf{H} = 
\setlength\arraycolsep{2pt}
\begin{bmatrix}
		 h[0] \!	&\!h[-1]& h[-2] 	& \ldots 	& \ldots &\!  h[-N+1] \\
		 h[1] \!	&\! h[0] & h[-1] 	& \ddots 	&   	 & \! \vdots 	 \\
		 h[2] \!	&\! h[1] 	& \ddots 	&  \ddots 	& \ddots &\!	\vdots	 \\
		 \vdots \!&\!\ddots & \ddots 	& \ddots	& h[-1] &\!  h[-2] 	 \\
		 \vdots \!& \! 		& \ddots 	& h[1] 		& h[0] 	 & \! h[-1]	 \\
		 h[N-1]&\!\ldots &\! \ldots 	& h[2] 		& h[1] 	 &\!  h[0]
\end{bmatrix},
 \end{equation}
 \normalsize where $\mathbf{h} =  [\, h[0], \ldots, h[L-2], h[L-1]\, ]^{\rm T}\in \mathbb{R}^{L \times 1}$ represents $L\leq N$ consecutive impulse response samples. For simplicity we assume causal LTI and finite impulse approximation $h[l] = 0$ for $l > L-1 ~\lor l < 0$ in this paper.

The sub-sampling matrix $\mathbf{B}$ is a wide matrix that characterizes the sampling scheme: 
\begin{equation}
\mathbf{B} = \bigoplus_{m=1}^M \bm{\kappa},\,  \in \{0,1\}^{M\times N},
\end{equation}
where $\kappa \in \{0,1\}^{1 \times R}$ such that:
\begin{equation*}
  \kappa[n] =
  \begin{cases}
    1, & \text{for}\ n = 1\\
    0, & \text{otherwise}
  \end{cases},
\end{equation*}
and  $\bigoplus$ ~denotes direct matrix sum. 
This matrix can be seen as containing a subset of the rows of an identity matrix, with all but each $R$'th row removed.

 The width of the matrices $\mathbf{B}$, $\mathbf{H}$ and $\mathbf{P}$ depends on how densely we represent the sampled signal after reconstruction. The matrix $\bPhi$ of width $N$ enables reconstruction of the input signal in the Nyquist-rate resolution. Moreover,  it is the minimum size, although a higher dimension may be chosen. An important factor is also the desired discrete representation accuracy of the filter's impulse response.  Here it is worth noticing that this RD framework considers low-pass filters but the literature also suggests the usage of an accumulate-and-dump architecture~\cite{Tropp:2010p3813, eldar2012compressed, becker2011}. In that case, an integrator with a reset system would be utilized~\cite{Tropp:2010p3813, becker2011, Mishali:2010p3825, Mishali:2010p4876}. From the transfer function perspective, the integrator behaves similar to a low-pass filter although it does not have flat pass-band response. Discussions regarding advantages and disadvantages of using one or the other solution are not the main concern of this article and  we recommend~\cite{becker2011} for more details.   
The notable difference in the case of using the accumulate-and-dump architecture is that it can be easily represented in a discrete model~\cite{eldar2012compressed}. Ideally, the integrator's impulse response is flat with unity amplitude and finite length of $R$. Using a low-pass filter, we deal with an infinite impulse response that needs to be approximated by the finite discrete-time model.

\subsection{Impulse response matrix} % (fold)
\label{sub:impulse_response_matrix}

An analog filter in the RD front-end can be described by a proper rational transfer function\cite{huelsman1993active}:
\begin{equation}
	\label{eq:transfer-funct-def}
	H_a(s) = {\displaystyle\sum_{b=0}^B{\lambda_b s^b }}\Biggl/{\displaystyle\sum^A_{b=0}{\beta_b s^b}}, 
\end{equation}
\noindent
where $\lambda_0, \beta_0, \ldots, \lambda_A, \beta_A  \in \mathbb{R}$, $B < A$, $s$ is the Laplace s-plane variable and $H_a(s)$ is the Laplace-transform of the impulse response $h_a(t)$.

In order to build the impulse response matrix $\mathbf{H}$, the essential task is to obtain the discrete impulse response of the analog filter which should accurately describe the filter. 
Many methods exist that transform the analog transfer function to the discrete-time counterpart e.g., bilinear transform (Tustin approximation) or the impulse invariance method\cite{oppenheim2010discrete}. The methods differ in mapping accuracy, computational complexity and filter type applicability. 

In cases where we deal with piecewise-constant frequency magnitude characteristics, such as lowpass, highpass and bandpass filters, the common approach is to use bilinear transformation\cite{oppenheim2010discrete}. The method essentially translates the filter transfer function (\ref{eq:transfer-funct-def}) from the continuous-time Laplace-domain to the discrete-time $z$-domain by the transformation: $s \leftarrow \frac{2}{T_{\text{z}}}\frac{z-1}{z+1},$ where
$z = \exp \left[j\omega T_{\text z}\right]$ and $T_{\text{z}}$ is the sampling period.

The discrete-time transfer function is expressed as follows:

\begin{equation}
	\label{eq:transfer-funct-Z2}
 	H_d(z) = \frac{\displaystyle\sum^B_{\ell=0}{b_\ell z^{-\ell} }}{\displaystyle\sum^A_{\ell=0}{a_\ell z^{-\ell}}}= \frac{b_0\displaystyle\prod^B_{\ell=1}{(1 -  d_{\ell}z^{-1} )}}{a_0\displaystyle\prod^A_{\ell=1}{(1- q_{\ell}z^{-1})}},
 \end{equation} 
\noindent
where $d_{\ell}$'s are the non-zero zeros of $H_d(z)$ and the $q_{\ell}$'s are the non-zero poles of $H_d(z)$. The discrete impulse response of the filter can be extracted through partial fraction expansion of $H_{\text{d}}$ \cite{oppenheim2010discrete}. Assuming that the poles are $1^{\text{st}}$ order, the transfer function can then be expressed as partial fractions
\cite{oppenheim2010discrete}:
\begin{equation}
	\label{eq:H-partial-frac}
	H_d(z) = \displaystyle\sum^A_{\ell=1} \frac{U_{\ell}}{1-q_{\ell}z^{-1}},
\end{equation}
where $U_\ell = (1-q_{\ell}z^{-1})H_d(z)|_{z = q_{\ell}} $.

The inverse $z$-transform is then calculated as a sum of partial inverse transforms, yielding discrete-time impulse response $h[1],\ldots,h[l]$. 

% subsection (end)

\subsection{Perturbed models} % (fold)
\label{sec:perturbed_models}
% section perturbed_models (end)
In CS analog acquisition, the inevitable situation, when the sampling front-end deviates from the initial model due to hardware imperfections, has been identified as measurement matrix perturbation. Furthermore, when the perturbation has a certain structure, we refer to it as structured perturbation of $\bPhi$~\cite{Herman2010}. When an additive noise in the compressed measurements is additionally included, we consider the model completely perturbed~\cite{Herman2010}. The error introduced by the sampling hardware results in an error
that is correlated with the input signal $x(t)$~\cite{Rosenbaum2010}. 

\begin{figure}[htb]
  \centering
  \includegraphics[width=0.82\columnwidth]{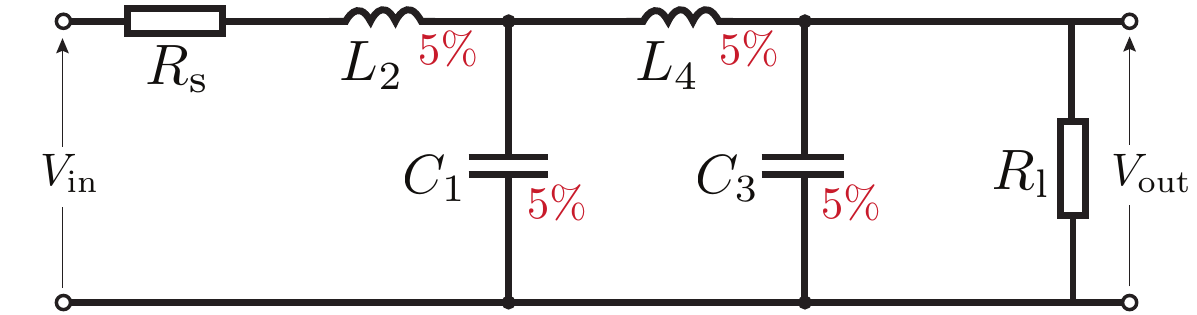}
  \caption{Example of a low-pass filter architecture (a 4th order
    double-resistively terminated LC network) used in the case study.}
  \label{fig:LCnet}
\end{figure}

\begin{figure}[tb]
  \setlength\figureheight{2.9cm}
  \setlength\figurewidth{7.5cm}
  \input{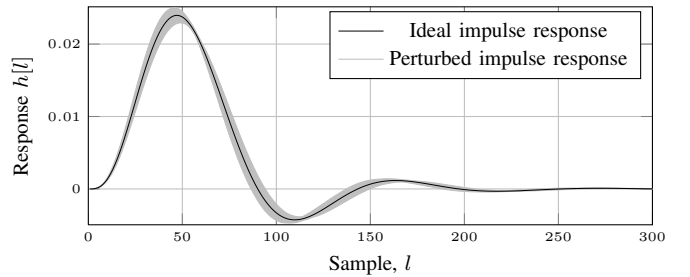} 
  \caption{Discrete impulse responses of the 4th order Butterworth
    low-pass filter from Fig.~\ref{fig:LCnet}. Ideal impulse response
    and 1000 randomly generated perturbed impulse responses from
    components subject to up to $\pm5$\% deviations.}
  \label{fig:figures_c16_v_ideal_butter_4th_500fc_impulse}
\end{figure}

One of the main sources of perturbation in $\bPhi$ is the low-pass filter and the sensitivity of the filter transfer function to non-exact component values~\cite{pankiewicz11}. Depending on the filter type, components might deviate from their nominal values due to the manufacturing process (component tolerance), device mismatch~\cite{kinget1996, steyaert1997,kinget2005} or parasitic components in the circuitry. These differences in component values change the shape of the implemented frequency response and cannot be controlled by the designer~\cite{Filanovsky:2012be}.

 Fig.~\ref{fig:figures_c16_v_ideal_butter_4th_500fc_impulse} shows the impulse responses of a double-resistively terminated LC network (Fig.~\ref{fig:LCnet}) realization of a 4th order low-pass Butterworth filter where the components differ up to 5\% from their nominal values according to a truncated (5\% of the mean) Gaussian distribution.

The error in the digital model of the impulse response can be considered additive  and  we can thus model the impulse response error matrix using the same structure as the $\mathbf{H}$ matrix has: 
\begin{equation}
	\label{eq:impulse_error}
	\widehat{\mathbf{H}} = \mathbf{H} + \mathbf{E},
\end{equation}
\noindent
where $\mathbf{H}$ represents the ideal impulse response matrix, $\widehat{\mathbf{H}}$  is the realized impulse response matrix, and $\mathbf{E}$ is the error matrix. The error matrix can be represented by the simplified\footnote{The model does not consider truncation of the impulse response of the filter which in theory is infinitely long. The error matrix is simplified to reflect a causal system.} structured model below:
    \small
\begin{equation}
    \mathbf{E} =
\begin{bmatrix}
		 e[0] 	& 0 &    0	& \ldots 	& \ldots &   0 \\
		 e[1] 	& e[0] &  0	& \ddots 	&   	 &  \vdots 	 \\
		 e[2] 	& e[1] 	& \ddots 	&  \ddots 	& \ddots &	\vdots	 \\
		 \vdots &\ddots & \ddots 	& \ddots	& 0 &   0	 \\
		 \vdots &  		& \ddots 	& e[1] 		& e[0] 	 &  0	 \\
		 e[N-1]&\ldots & \ldots 	& e[2] 		& e[1] 	 &  e[0]
\end{bmatrix},
\label{eq:E-matrix}
 \end{equation}
 \normalsize
where $\mathbf{e} = [e[0],\ldots,e[L-1]]^{\mathrm{T}}\in\mathbb{R}^{L\times1}$ and $L \leq N$ represents the error vector between the actual impulse response of the system $\mathbf{\hat{h}}$ and the modeled $\mathbf{h}$:
\begin{equation}
 	\label{eq:init_error}
 	\mathbf{e} = \hat{\mathbf{h}} - \mathbf{h}.
 \end{equation} 

% subsection measurement_matrix (end)
Previous research in~\cite{pankiewicz11} has shown that this kind of structured perturbation degrades the reconstructed signal quality of generic CS reconstruction algorithms. Most of the generic reconstruction algorithms can only deal with partially perturbed models $\y = \bPhi \x + \mathbf{w}$, considering added noise $\mathbf{w}$ to the measurements\cite{Candes2006c, Donoho:2006ja, Tropp:2006vb}. The reconstruction algorithms assume the nominal hardware component values represented by the discrete measurement model $\y = \bPhi\x$. In practice, the hardware system performs the sampling operation $\hat{\y} = \hat{\bPhi}\x$, where $\hat{\bPhi} = \mathbf{B\widehat{H}}\mathbf{P} = \bPhi + \mathbf{BEP}$. Consequently, the measurements $\mathbf{\hat{y}}$ obtained from the hardware system correspond to the ideal measurements with correlated additive noise $\mathbf{\hat{y}} = \y + \mathbf{BEPx}$.

\section{Calibration methodology} % (fold)
\label{sec:calibration_methodology}
In the calibration scenario proposed, we exploit the structure of the perturbation $\mathbf{E}$ to perform supervised calibration utilizing only $M_{\rm q} \sim L$ samples of a known arbitrary sequence. 

\subsection{Linear estimation of the impulse response model error } % (fold)
\label{sub:linear_estimation}
In order to calibrate the existing measurement matrix, we need to estimate the error matrix $\mathbf{E}$. Assuming that the RD system can sample a known signal
$\x_{\rm q} \in \mathbb{R}^{S\times 1}$, $ M_{\rm q} < S$ (e.g. from a generator), we can exploit the fact that the structure of the error is already known:

\begin{equation}
	\label{eq:y_test}
	\hat \y_{\rm q} = \bm{\hat{\Phi}} \x_{\rm q}.
\end{equation}

\noindent Under the assumption of known input signal $\x_{\rm q}$ and front-end model $\bPhi$, we can rearrange the measurement equation such that $\mathbf{E}$ becomes the only unknown:
\begin{equation}
  \label{eq:perturebd_measurements_extended1}
  \hat{\y}_{\rm q} = \bPhi\x_{\rm q} + \mathbf{BEP}\x_{\rm q}.
\end{equation}

\noindent Given $\y_{\rm q} \in \mathbb{R}^{M_q\times 1} $ and the ideal measurements $\y_{\rm q} = \bPhi \x_{\rm q}$, the difference between the modeled low-pass filter response and the actual response can be modeled as:
\begin{equation}
  \label{eq:perturebd_measurements_pre-ext}
  \hat{\y}_{\rm q} - \y_{\rm q} = \mathbf{BEP}\x_{\rm q}.
\end{equation}

\noindent Furthermore, the roles of $\mathbf{E}~\text{and}~\x_{\rm q}$ can be interchanged as follows: 
\begin{equation}
  \label{eq:interchange}
  \mathbf{BEP} \x_{\rm q} = \mathbf{De}
\end{equation}
 where:
\begin{equation}
\label{eq:D-matrix}
  \mathbf{D} = 
  \begin{bmatrix}
 d[1] &  \ldots & d[L]    \\ 
 d[R+1] &   \ldots   &  d[R+L]  \\ 
 d[2R+1] &   \ldots   &  d[2R+L]  \\ 
\vdots &   & \ldots  \\
 d[N-L+1] &  \ldots   &  d[N]
\end{bmatrix}\in \mathbb{R}^{M\times L}.
\end{equation}
The matrix $\mathbf{D}$ is based on \eqref{eq:demodulation-digital} and $L$ is the size of the discrete impulse response vector. When $L > R$, $\mathbf{D}$ becomes rank deficient and should be tailored by truncating its first $L/R$ rows  and discarding the first $L/R$ measurements of $\mathbf{\hat y_{q}}$  and $ \mathbf{y_{q}}$.
To avoid rounding errors, $L/R \in \mathbb{Z}$ is required.

Equation (\ref{eq:perturebd_measurements_pre-ext}) can be further rewritten using (\ref{eq:interchange}) to the following form:
\begin{equation}
	\label{eq:e_estimate}
	\mathbf{D\,e} = \mathbf{\hat y_{q}} - \mathbf{y_{q}}.
\end{equation}

If the system (\ref{eq:e_estimate}) is overdetermined ($M > L$), we can use a least-squares estimator to calculate  $\mathbf{e}$. Due to the banded-Toeplitz structure of $\mathbf{H}$ and $\mathbf{E}$, estimating $\mathbf{e}$ from (\ref{eq:perturebd_measurements_pre-ext}) consequently amounts to calibrating the entire $\hat{\bPhi} = \mathbf{B\widehat{H}}\mathbf{P}$ matrix in the reconstruction. The process can be defined as:
\begin{equation}
  \label{eq:least-squares}
  \underset{{\mathbf{e} \in \reals^{L\times1}}}{\text{minimize}}~ \|\mathbf{De} - \mathbf{\check{y}}\|_{2}^2,
  \end{equation}

\noindent where:
\begin{equation}
\mathbf{\check{y}} =  \y_{\rm q} - \hat{\y}_{\rm q}.
\label{eq:y_diff}
\end{equation}
The impulse response model error can also be estimated  in cases where the number of calibration measurements $M \leq L$. In this case we could use Tikhonov-regularized least-squares defined as follows\cite{tarantola2005inverse}:

 \begin{equation}
 \begin{array}{ll}\label{eq:least-squares-TK}
 \displaystyle \mathop{\mbox{minimize}}_{\mathbf{e} \in \reals^{L\times1}} & \|\mathbf{De} - \mathbf{\check{y}}\|_{2}^2\\
 \mbox{subject to} &  \|\mathbf{Ge}\|_{2}^2 \leq \gamma, \\
 \end{array}
\end{equation}

\noindent where $\gamma > 0$ is a regularization parameter determining the sensitivity of the solution and $\mathbf{G}\in\mathbb{R}^{L\times L}$ is a regularization operator. This particular problem has been posed with $\gamma = \lambda_{\min(\mathbf{D}\mathbf{D^{\rm{T}}})}$ and $\mathbf{G} = {\rm diag}\left(\mathbf{g},\mathbf{v}\right)$, where $\mathbf{g}\in\{1\}^{L/2\times 1}$, $\mathbf{v}\in\{0\}^{L/2\times 1}$  .

Based on (\ref{eq:least-squares}) and~(\ref{eq:least-squares-TK}) we propose the model-based calibration (MBC) algorithm for the RD framework presented in Algorithm~\ref{alg1}.

\begin{algorithm}[htb!]
  \algsetup{linenosize=\footnotesize,linenodelimiter=.}
   \caption{Model-based calibration (MBC) of the impulse response model $\mathbf{\widehat{H}}$ in the random demodulation architecture.}
  \label{alg1}
\begin{algorithmic}[1]
     \REQUIRE  known signal $\x_{\rm q} \in \mathbb{R}^{S\times 1}$, chipping sequence $p[1],\ldots,p[N]$, number of measurements: \Mq, initial impulse response size $L$,  $\bPhi \in \mathbb{R}^{M_{\text{q}}\times N}$.
    \STATE $\hat{\y}_{\rm q} \gets$  $\widehat{\bPhi}\x_{\rm q}$ $\quad \hookleftarrow$ \eqref{eq:y_test}
    \STATE $\y_{\rm q} \leftarrow \bPhi \x_{{\rm q}}$ $\quad \hookleftarrow$ \eqref{eq:cs_primer}
    \STATE $\mathbf{\check{y}} \leftarrow \y_{\rm q} - \hat{\y}_{\rm q}$ $\quad \hookleftarrow$ \eqref{eq:y_diff}
    \STATE $\mathbf{D}[1:M,1:L] \leftarrow$  $d[1],\ldots,d[N]$ $\quad \hookleftarrow$  \eqref{eq:demodulation-digital}, \eqref{eq:D-matrix}
    \IF {$L > R$}
    \STATE $\mathbf{D} \leftarrow \mathbf{D}[\frac{L}{R}:\text{end},:] \land \frac{L}{R} \in \mathbb{N}$
    \STATE $\check{\y} \leftarrow \check{\y}[\frac{L}{R}:\text{end},:]$
    \STATE $M_{\text{q}} \leftarrow M_{\text{q}} -  \frac{L}{R}$
    \ENDIF
	\IF{ \Mq $\geq L$}
	\STATE calculate $\hat{\mathbf{e}}$ using  \eqref{eq:least-squares}
	\ELSE
	\STATE calculate $ \hat{\mathbf{e}}$ using  \eqref{eq:least-squares-TK}
	\ENDIF
    \STATE $\mathbf{\mathring{h}} \leftarrow \mathbf{h - \hat{e}}$
    \STATE $\mathbf{\mathring{H}} \leftarrow$ $\mathbf{\mathring{h}}$ $\quad \hookleftarrow$  \eqref{eq:H-matrix} 
    \STATE  $\bm{\mathbf{\mathring{\Phi}}} \leftarrow \mathbf{B}\mathbf{\mathring{H}}\mathbf{P}$ $\quad \hookleftarrow$ \eqref{eq:psi_extended}
    \RETURN $\bm{\mathbf{\mathring{\Phi}}}\in \mathbf{R}^{M\times N}$
\end{algorithmic}
\end{algorithm}

Having estimated $\mathbf{\hat{e}}$, from (\ref{eq:least-squares}) or (\ref{eq:least-squares-TK}), we can create  a calibrated impulse response matrix $\mathbf{\mathring{H}}$ and thus an updated measurement matrix $\boldsymbol{\mathring{\Phi}}$. The method enables calibration of the RD filter matrix without any additional changes in the architecture and it is compatible with arbitrary CS reconstruction algorithms.
Even though the method in principle calibrates the impulse response $\mathbf{h}$, it can compensate for more than only filter model perturbation. The uncertainty of e.g., an amplifier gain can be calibrated, where $\mathbf{\hat{y}_{\rm q} = \bm{\hat \Phi}x_{q}}$, as long as we deal with an uncertainty of a linear system.

\section{Simulation framework} 
\label{sec:simulation_framework}
We design a set of numerical simulations to verify and evaluate the proposed calibration approaches. The simulation environment\footnote{To reproduce the experiments -- the MATLAB code is freely available at: \url{http://www.sparsesampling.com/mbc}}, developed in MATLAB 2012a and executed on PC, Ubuntu 12.04 LTS--Intel X5670  \@2.93 GHz, is divided in two separate parts:
\begin{enumerate}
 	\item Modeling component deviations in the low-pass filter according to  specified  tolerances and evaluating RD performance under perturbed models without calibration. 
 	\item Performance analysis of the calibration Algorithm~\ref{alg1}. The analysis is based on two Monte Carlo simulation schemes. The first scheme evaluates  the
error between the calibrated impulse response and its original.  
value. The second approach focuses on the BPDN reconstruction with calibrated measurement matrix.   
 \end{enumerate} 

\vspace{-0.2cm}
\subsection{Filter case study} % (fold)
\label{sub:example_filter_case_study_}
We consider a passive low-pass filter architecture utilized by  the RD front-end. One of the main drawbacks in using passive filters is their transfer function sensitivity to element (component) changes. For our experiments we have chosen the doubly resistive terminated LC ladder network designed for maximum power transfer and therefore with superior sensitivity properties. The passive architecture has been chosen here to facilitate modeling of the components variations, but the observations do apply to any discrepancy between filter model and the actual hardware.

The filter in Fig.~\ref{fig:LCnet} can be characterized by the transfer function:

\begin{gather}
  \label{eq:lc-transfer}
  H_{\rm{LC}}(s) = \frac{\lambda_0}{\displaystyle\sum^4_{c=0}{\beta_c s^c}},
  \intertext{where}
  \begin{aligned}
     \beta_0 &= \frac{\Rs}{\Rl} + 1,~~\beta_1 = L_{4} + L_{2} + C_{1}\frac{\Rs}{\Rl} + C_{3}\frac{\Rs}{\Rl}, \\
    \beta_2 & = L_{4}C_{1} + L_{2}C_{1}+ L_{2}C_{3}\frac{\Rs}{\Rl} + L_{4}C_{3}, \\
    \beta_3 &= L_{4}L_{2}C_{3}+L_{2}C_{3}C_{1}\frac{\Rs}{\Rl},\\
    \beta_4 & = L_{4}L_{2}C_{3}C_{4}, ~~\lambda_0 = \sqrt{\frac{4\Rs}{\Rl}}.\\
  \end{aligned}\notag
\end{gather}

We conduct a series of numerical experiments to evaluate the effect of filter component deviations on the CS reconstruction quality. The test strategy is divided into four scenarios. Each scenario considers deviation in one filter component $\{C_1,~C_3,~L_2~\rm{or}~L_4\}$, affecting the filter characteristics, and therefore causing measurement matrix mismatch during the reconstruction stage. This allows us to investigate how much a single component variation can influence the reconstruction.
 Furthermore, we consider deviation of all components and apply the proposed calibration approach in~(\ref{eq:least-squares}) or~(\ref{eq:least-squares-TK}) to compensate the filter imperfections and evaluate the performance. 
 In our experiments, we have used a multi-tone signal with $K-1$ randomly chosen tones from a tone dictionary $F\in \left\{ 2,3, \ldots ,1500 \right\}~\si{\hertz}$ and an amplitude dictionary $a\in \left\{1,2, \ldots ,10  \right\}$. We have used a calibrating signal $\x_{\text{q}}$ with $K=10$ tones and  input signal $\x$ for the RD reconstruction tests with $K=5$. One tone is  always set to $1500$~\si{\hertz} to provide consistent Nyquist frequency $f_{2\mathcal{B}}$ . In the reconstruction stage, the framework  processes 1\,\si{\second} of an input signal $\x$, which is represented by $N = 12600$ samples $(f_{N} = 4.2 \cdot 2\mathcal{B})$; the oversampled representation is used to emulate continous-time analog signals. We have synthesized the low-pass filter with a $3$~\si{\decibel} cut-off frequency $f_{\rm c} = 500~\si{\hertz}$ as Butterworth and Chebyshev approximations with the  component values listed in Table~\ref{tab:components}~\cite{huelsman1993active}.

\begin{table}[h!]
\centering
\renewcommand{\arraystretch}{1.3}
\caption{4th order LC-ladder components in considered approximations for $f_c = 500$~\si{\hertz}.}
\begin{tabular}{|c|c|c|c|c|c|c|}
\hline
			& $C_{1}$ & $C_{3}$ & $L_{2}$ & $L_{4}$ & $R_{s}$ & $R_{l}$ \\
		& $[\si{\micro\farad}]$ & $[\si{\micro\farad}]$ & $[\si{\milli\henry}]$ & $[\si{\milli\henry}]$ & $[\si{\ohm}]$ & $[\si{\ohm}]$ \\
			\hline	
Butterworth & \num[
         round-mode = places,add-decimal-zero = true, add-integer-zero = true,
         round-precision = 1]{4.8725} & \num[
         round-mode = places,add-decimal-zero = true, add-integer-zero = true,
         round-precision = 1]{11.7632} & \num[
         round-mode = places,add-decimal-zero = true, add-integer-zero = true,
         round-precision = 1]{29.408}& \num[
         round-mode = places,add-decimal-zero = true, add-integer-zero = true,
         round-precision = 1]{12.1812} &  \num[
         round-mode = places,add-decimal-zero = true, add-integer-zero = true,
         round-precision = 1]{50.} & \num[
         round-mode = places,add-decimal-zero = true, add-integer-zero = true,
         round-precision = 1]{50.}\\
Chebyshev   & \num[round-mode = places,add-decimal-zero = true, add-integer-zero = true, round-precision = 1]
{5.7812} & \num[round-mode = places,add-decimal-zero = true, add-integer-zero = true, round-precision = 1]
{7.9132} & \num[round-mode = places,add-decimal-zero = true, add-integer-zero = true, round-precision = 1]
{36.0591} & \num[round-mode = places,add-decimal-zero = true, add-integer-zero = true, round-precision = 1]
{24.6173} & \num[round-mode = places,add-decimal-zero = true, add-integer-zero = true, round-precision = 1]
{50.} & \num[round-mode = places,add-decimal-zero = true, add-integer-zero = true, round-precision = 1]
{100.}\\
\hline
\end{tabular}
\label{tab:components}
\end{table}

The transfer function $H_{\rm LC}(s)$ has been discretized using bilinear transform and a sampling frequency $(\frac{1}{T_{\text{z}}})$ of 13~\si{\kilo\hertz}. The calculated discrete impulse response $h_{\rm LC}[l]$ has been used to define the measurement matrix $\bPhi \in \mathbb{R}^{M\times N}$. The sub-sampling frequency $f_{\rm s}$ has been set to 1.05~\si{\kilo\hertz} ($2 f_{\rm cut}$). Using IDFT as dictionary $\bPsi \in \mathbb{C}^{N \times N}$ in the reconstruction algorithm enables reconstruction $(\hat{\x})$ of the input signals $\x \in \mathbb{R}^{N \times 1}$.
The reconstruction quality is assessed in terms of the Signal-to-Noise Ratio $(\rm{SNR})$ defined as:
\begin{equation}
 	\label{eq:SNR}
 	\xi = 20\log_{10}\left(\frac{\|\x\|_{2}}{\|\x - \hat{\x}\|_{2}}\right).
 \end{equation}  

 We have assumed a production line yielding components according to the following expression~\cite{oakland2003statistical}:

\begin{equation}
\label{eq:truncg}
\theta_{\mu, \sigma} (c) = \left\{ \begin{array}{@{}ll@{}}
 \frac{\displaystyle 1}{{\displaystyle \sigma \sqrt {2\pi } }}\displaystyle \exp\!\left[ -\frac{(c-\mu)^2}{2\sigma^2} \right] ,&\mbox{for $\vert c-\mu \vert \leq \sigma$}
\\
 \\
 0, & \text{otherwise}\\
  \end{array}.\right.
\end{equation}

\noindent In this article we refer to  \eqref{eq:truncg} as a truncated Gaussian distribution. The standard deviation $\sigma$ is set to $2\%$ of the  nominal component value $\mu$. Additional quality control with an aim of $\max$ (e.g., $2\%$) tolerance is modeled as a truncation of the component distribution.

In the initial experiment we have performed Monte Carlo simulations, analyzing the effect of single-component deviation. The simulations considered 1000 different component values according to \eqref{eq:truncg} for considered single-component variation of $L_2, L_4, C_1,\text{~and~} C4$ in the Butterworth low-pass filter. Reconstruction was performed with the BPDN algorithm using the SPGL1\footnote{A solver for large-scale sparse reconstruction \url{http://www.cs.ubc.ca/labs/scl/spgl1}.} solver\cite{BergFriedlander:2008, Chen2001}. The results are presented in Fig.~\ref{fig:gauss}. Despite using the least sensitive passive filter architecture, the single-component deviation according to \eqref{eq:truncg} with $\mu \in \{L_2, L_4, C_1, C_3\}$, and $\sigma=0.02\mu$, causes the average reconstruction quality of approximately 47--54~\si{\deci\bel} as opposed to 87~\si{\deci\bel} in case of a known model. A small reconstruction variation in the known model case is caused by the change of filter characteristics due to component variation but it is relatively small compared to the unknown perturbation case. 

\begin{figure}[ht!]  
    \setlength\figureheight{1.3cm}
    \setlength\figurewidth{3.2cm}
 % This file was created by matlab2tikz v0.3.0.
% Copyright (c) 2008--2012, Nico Schlรถmer <nico.schloemer@gmail.com>
% All rights reserved.
% 
% The latest updates can be retrieved from
%   http://www.mathworks.com/matlabcentral/fileexchange/22022-matlab2tikz
% where you can also make suggestions and rate matlab2tikz.
% 
% 
% 
\footnotesize

\begin{tikzpicture}
\begin{axis}[%
 /pgf/number format/.cd,
 		use period,
        1000 sep={},
width=\figurewidth,
height=\figureheight,
scale only axis,
xmin=37.4898800697399, xmax=89.3441198633178,
ymin=0, ymax=274,
ylabel={Instances},
name=plot3,
legend style={anchor=north east,draw=none,fill=none,align=right}, empty legend
]
\addplot[ybar,bar width=0.0624999999999998\figurewidth,fill=red,draw=black] plot coordinates{ (39.1103250632892,274) (42.3512150503878,192) (45.5921050374864,146) (48.8329950245851,99) (52.0738850116837,89) (55.3147749987823,61) (58.5556649858809,37) (61.7965549729795,44) (65.0374449600782,16) (68.2783349471768,7) (71.5192249342754,6) (74.760114921374,7) (78.0010049084726,7) (81.2418948955712,6) (84.4827848826699,2) (87.7236748697685,7) };

%\addlegendimage{empty legend}
\addlegendentry[text width=40pt,text depth=]{{\text{C3,~}$\mu\text{=47.4}$} {$\sigma\text{=9.4}$}};

\addplot [
color=black,
solid,
forget plot
]
coordinates{
 (37.4898800697399,0)(89.3441198633178,0) 
};
\end{axis}

\begin{axis}[%
width=\figurewidth,
height=\figureheight,
scale only axis,
xmin=37.4898800697399, xmax=89.3441198633178,
ymin=0, ymax=227,
ylabel={Instances},
name=plot1,
at=(plot3.above north west), anchor=below south west,
legend style={anchor=north east,draw=none,fill=none,align=right}, empty legend
]
\addplot[ybar,bar width=0.0624999999999998\figurewidth,fill=red,draw=black] plot coordinates{ (39.1103250632892,0) (42.3512150503878,0) (45.5921050374864,227) (48.8329950245851,212) (52.0738850116837,175) (55.3147749987823,103) (58.5556649858809,86) (61.7965549729795,49) (65.0374449600782,38) (68.2783349471768,35) (71.5192249342754,27) (74.760114921374,12) (78.0010049084726,14) (81.2418948955712,8) (84.4827848826699,11) (87.7236748697685,3) };

\addlegendentry{{\text{C1,~}$\mu\text{=54.2}$}\\ {$\sigma\text{=8.9}$}};

\addplot [
color=black,
solid,
forget plot
]
coordinates{
 (37.4898800697399,0)(89.3441198633178,0) 
};
\end{axis}

\begin{axis}[%
 /pgf/number format/.cd,
	set thousands separator={},
	set decimal separator={.},
	width=\figurewidth,
height=\figureheight,
scale only axis,
xmin=87.897841945916, xmax=88.3182672439968,
xtick = {87.95, 88.1, 88.2, 88.3},
ymin=0, ymax=950,
ytick = {0, 500, 950},
name=plot2,
at=(plot1.right of south east), anchor=left of south west,
legend style={anchor=north east,draw=none,fill=none,align=right}, empty legend
]
\addplot[ybar,bar width=0.0624999999999873\figurewidth,fill=blue,draw=black] plot coordinates{ (87.910980236481,978) (87.937256817611,13) (87.9635333987411,9) (87.9898099798711,0) (88.0160865610012,0) (88.0423631421313,0) (88.0686397232613,0) (88.0949163043914,0) (88.1211928855214,0) (88.1474694666515,0) (88.1737460477815,0) (88.2000226289116,0) (88.2262992100416,0) (88.2525757911717,0) (88.2788523723017,0) (88.3051289534318,0) };

\addlegendentry{{$\text{C1,~}\mu\text{=87.6}$}\\{$\sigma\text{=0.2}$}};

\addplot [
color=black,
solid,
forget plot
]
coordinates{
 (87.897841945916,0)(88.3182672439968,0) 
};
\end{axis}

\begin{axis}[%
 /pgf/number format/.cd,
	set thousands separator={},
	set decimal separator={.},
width=\figurewidth,
height=\figureheight,
scale only axis,
xmin=87.897841945916, xmax=88.3182672439968,
xtick = {87.95, 88.1, 88.2, 88.3},
ymin=0, ymax=800,
ytick = {0, 400, 800},
name=plot4,
at=(plot2.below south west), anchor=above north west,
legend style={anchor=north east,draw=none,fill=none,align=right}, empty legend
]
\addplot[ybar,bar width=0.0624999999999873\figurewidth,fill=blue,draw=black] plot coordinates{ (87.910980236481,737) (87.937256817611,19) (87.9635333987411,23) (87.9898099798711,13) (88.0160865610012,7) (88.0423631421313,7) (88.0686397232613,12) (88.0949163043914,20) (88.1211928855214,31) (88.1474694666515,26) (88.1737460477815,23) (88.2000226289116,31) (88.2262992100416,28) (88.2525757911717,12) (88.2788523723017,7) (88.3051289534318,4) };

\addlegendentry{{$\text{C3,~}\mu\text{=87.5}
$}\\{$\sigma\text{=0.6}$}};

\addplot [
color=black,
solid,
forget plot
]
coordinates{
 (87.897841945916,0)(88.3182672439968,0) 
};
\end{axis}

\begin{axis}[%
 /pgf/number format/.cd,
	set thousands separator={},
	set decimal separator={.},
width=\figurewidth,
height=\figureheight,
scale only axis,
xmin=87.897841945916, xmax=88.3182672439968,
xtick = {87.95, 88.1, 88.2, 88.3},
ymin=0, ymax=800,
ytick = {0, 400, 800},
name=plot6,
at=(plot4.below south west), anchor=above north west,
legend style={anchor=north east,draw=none,fill=none,align=right}, empty legend
]
\addplot[ybar,bar width=0.0624999999999873\figurewidth,fill=blue,draw=black] plot coordinates{ (87.910980236481,744) (87.937256817611,30) (87.9635333987411,29) (87.9898099798711,11) (88.0160865610012,9) (88.0423631421313,10) (88.0686397232613,13) (88.0949163043914,17) (88.1211928855214,29) (88.1474694666515,18) (88.1737460477815,25) (88.2000226289116,22) (88.2262992100416,25) (88.2525757911717,11) (88.2788523723017,7) (88.3051289534318,0) };

\addlegendentry{{$\text{L2,~}\mu\text{=87.4}$}\\{$\sigma\text{=0.6}$}};

\addplot [
color=black,
solid,
forget plot
]
coordinates{
 (87.897841945916,0)(88.3182672439968,0) 
};
\end{axis}

\begin{axis}[%
width=\figurewidth,
height=\figureheight,
scale only axis,
xmin=37.4898800697399, xmax=89.3441198633178,
ymin=0, ymax=268,
ylabel={Instances},
name=plot5,
at=(plot6.left of south west), anchor=right of south east,
legend style={anchor=north east,draw=none,fill=none,align=right}, empty legend
]
\addplot[ybar,bar width=0.0624999999999998\figurewidth,fill=red,draw=black] plot coordinates{ (39.1103250632892,268) (42.3512150503878,184) (45.5921050374864,158) (48.8329950245851,105) (52.0738850116837,86) (55.3147749987823,54) (58.5556649858809,45) (61.7965549729795,40) (65.0374449600782,20) (68.2783349471768,6) (71.5192249342754,14) (74.760114921374,7) (78.0010049084726,7) (81.2418948955712,4) (84.4827848826699,1) (87.7236748697685,1) };

\addlegendentry{{$\text{L2,~}\mu\text{=47.4}$}\\{$\sigma\text{=9.0}$}};

\addplot [
color=black,
solid,
forget plot
]
coordinates{
 (37.4898800697399,0)(89.3441198633178,0) 
};
\end{axis}

\begin{axis}[%
width=\figurewidth,
height=\figureheight,
scale only axis,
xmin=37.4898800697399, xmax=89.3441198633178,
xlabel={SNR [dB]},
ymin=0, ymax=227,
ylabel={Instances},
name=plot7,
at=(plot5.below south west), anchor=above north west,
legend style={anchor=north east,draw=none,fill=none,align=right}, empty legend
]
\addplot[ybar,bar width=0.0624999999999998\figurewidth,fill=red,draw=black] plot coordinates{ (39.1103250632892,0) (42.3512150503878,0) (45.5921050374864,227) (48.8329950245851,212) (52.0738850116837,175) (55.3147749987823,103) (58.5556649858809,86) (61.7965549729795,49) (65.0374449600782,38) (68.2783349471768,35) (71.5192249342754,27) (74.760114921374,12) (78.0010049084726,14) (81.2418948955712,8) (84.4827848826699,11) (87.7236748697685,3) };

\addlegendentry{{$\text{L4,~}\mu\text{=54.2}$}\\{$\sigma\text{=8.9}$}};

\addplot [
color=black,
solid,
forget plot
]
coordinates{
 (37.4898800697399,0)(89.3441198633178,0) 
};
\end{axis}

\begin{axis}[%
 /pgf/number format/.cd,
	set thousands separator={},
	set decimal separator={.},
width=\figurewidth,
height=\figureheight,
scale only axis,
xmin=87.897841945916, xmax=88.3182672439968,
xtick = {87.95, 88.1, 88.2, 88.3},
xlabel={SNR [dB]},
ymin=0, ymax=950,
ytick = {0, 500, 950},
at=(plot7.right of south east), anchor=left of south west,
legend style={anchor=north east,draw=none,fill=none,align=right}, empty legend
]
\addplot[ybar,bar width=0.0624999999999873\figurewidth,fill=blue,draw=black] plot coordinates{ (87.910980236481,978) (87.937256817611,13) (87.9635333987411,9) (87.9898099798711,0) (88.0160865610012,0) (88.0423631421313,0) (88.0686397232613,0) (88.0949163043914,0) (88.1211928855214,0) (88.1474694666515,0) (88.1737460477815,0) (88.2000226289116,0) (88.2262992100416,0) (88.2525757911717,0) (88.2788523723017,0) (88.3051289534318,0) };

\addlegendentry{{$\text{L4,~}\mu\text{=87.6}$}\\{$\sigma\text{=0.2}$}};

\addplot [
color=black,
solid,
forget plot
]
coordinates{
 (87.897841945916,0)(88.3182672439968,0) 
};
\end{axis}
\end{tikzpicture}%	
	\caption{Instances versus SNR bins for 1000 Monte Carlo simulations. Right column  represents BPDN reconstruction quality using the measurement matrix matching every instance of deviated filter. The left column corresponds to the reconstruction quality assuming ideal filter impulse response in the model.} 	
	\label{fig:gauss}
\end{figure}
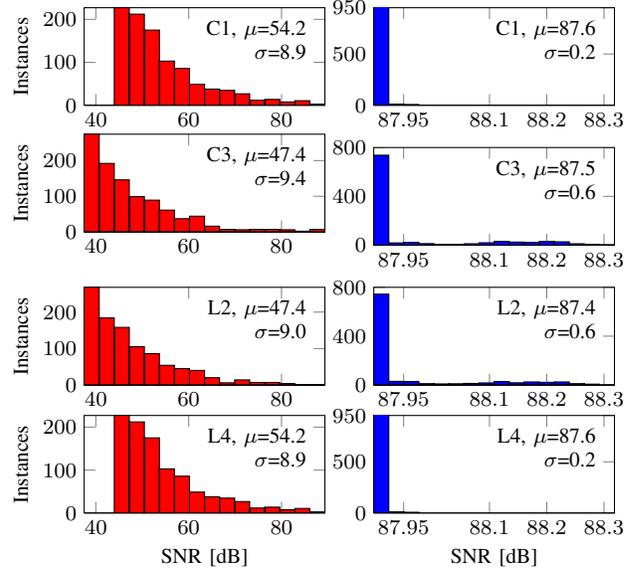

\subsection{Calibration performance evaluation} % (fold)
\label{sub:results}
 Algorithms~\ref{alg2} and \ref{alg3} provide a general overview of~the~MBC algorithm performance analysis. Algorithm~\ref{alg2} describes the process of obtaining compressed samples and evaluating reconstruction performance assuming no calibration has been done. 

 We have tested the described 4th order LC ladder network in which all the components were subject to nominal value deviations according to (\ref{eq:truncg}). The distribution of $p = 3000$ has been simulated, creating 3000 different component sets: $[\{L_{2,1}, L_{4,1}, C_{1,1}, C_{3,1}\}\ldots\{L_{2,p}, L_{4,p}, C_{1,p}, C_{3,p}\}]$. In each case we have calculated the Root Mean Square Error (RMSE) of \eqref{eq:init_error} -- which we denote $Q({\rm \mathbf{e}_{\rm p}})$. Further,  we have performed calibration according to Algorithm~\ref{alg1} and calculated the RMSE values between the calibrated impulse response $\mathbf{\mathring{h}}$  and the actual $\mathbf{\hat{h}}$ (deviating) impulse response:

\vspace{-0.15cm}
\begin{equation}
	\label{rmse_calibrated}
	Q({\rm \hat{\mathbf{e}}_{\rm p}}) = \frac{1}{\sqrt{L}}\|\mathbf{\mathring{h}} - \hat{\mathbf{h}}\|_2.
\end{equation}
   Two calibrating scenarios have been considered, where the first assumes taking   $M \geq L$ measurements of $\x_{\text{q}}$ and estimates the error in the impulse response by solving \eqref{eq:least-squares}. The second scenario examines an undetermined system, where $M < L$. The system is calibrated by solving~(\ref{eq:least-squares-TK}). 
The normalized results of the Monte Carlo simulations (with respect to iteration numbers) of $Q({\rm \mathbf{e}_{\rm p}})$  and  $Q({\rm \hat{\mathbf{e}}_{\rm p}})$ are shown in Fig.~\ref{MC_histograms}. 

\begin{algorithm}[htb!]
  \caption{RD performance under filter perturbations}
  \label{alg2}
  \algsetup{linenosize=\footnotesize,linenodelimiter=.}
  \begin{algorithmic}[1]
    \REQUIRE $\x \in \mathbb{R}^{N \times 1}$, component tolerances, CS measurements, sub-sampling frequency $f_{\rm s}$, $p$, $f_{\rm c} = \frac{f_{\rm s}}{2}$, $M$, $N$
    \STATE synthesize filter \eqref{eq:lc-transfer} $\hookleftarrow \mu_0=\{L_{2,0}, L_{4,0}, C_{1,0}, C_{3,0}\}$ according to (Table~\ref{tab:components}); obtain: $\textbf{H}\in\mathbb R^{M\times N} $ 
    \STATE generate chipping sequence $p(n) $ and the matrix in \eqref{eq:chip_seq_d}
    \STATE construct $\bPhi \in \mathbb{R}^{M \times N}$ according to \eqref{eq:psi_extended}
    \STATE create sparsity basis IDFT matrix  $ \bPsi \in \mathbb{C}^{N\times N}$ 
    \FOR{$c=1 ~to~ {\text{p}}$}
    \STATE  $\mu_c = \{L_{2,c}, L_{4,c}, C_{1,c}, C_{3,c}\}$ $\quad \hookleftarrow$ \eqref{eq:truncg}
    \STATE sample $ \mathbf{\hat y_c} = \bm\hat{\Phi}_c\x $ analogous to \eqref{eq:y_test}    
    \STATE reconstruct $\hat\balpha_c \gets$ SPGL1($\bPhi$, $\bPsi$, ${\hat \y}$) $\quad \hookleftarrow$  \eqref{eq:BPDN}
    \STATE recover $\x$: $\hat{\x}_c = \Re\{\bPsi\hat\balpha_c\}$, $\Re$ denotes real-part
    \STATE $Q({\rm \mathbf{e}}_{c}) = \frac{1}{\sqrt{L}}\|\mathbf{h} - \hat{\mathbf{h}}_c\|_2$
    \STATE  Compute SNR $\xi_c  \quad \hookleftarrow$ \eqref{eq:SNR}
    \ENDFOR
    \RETURN Performance merits: $Q(\mathbf{e}_c)$ (RMSE), $\xi_{\text{c}}$[dB]
  \end{algorithmic} 
\end{algorithm}

\begin{algorithm}[htb!]
  \caption{MBC performance analysis}
  \label{alg3} 
  \algsetup{linenosize=\footnotesize,linenodelimiter=.}
  \begin{algorithmic}[1]
    \REQUIRE reuse data from Algorithm~\ref{alg2} ($\x$, $\hat{\y}$, $\bPhi$,$\mu_c$)
    \FOR{$c=1\ {\bf to}\ {\text{p}}$}
    \STATE obtain $\bm{\mathring{\Phi}_c}$ using Algorithm~\ref{alg1}
   \STATE reconstruct $\hat\balpha_c$ \eqref{eq:BPDN}
    using SPGL1 with $\bm{\mathring{\Phi}_c}$, $\bPsi$
    \STATE recover $\x$: $\hat{\x}_c = \Re\{\bPsi\hat\balpha_c\} $
    \STATE $Q(\mathbf{\hat{e}}_c) \quad \hookleftarrow$ \eqref{rmse_calibrated} 
    \STATE  Compute SNR $\mathring{\xi}_c \quad \hookleftarrow$ \eqref{eq:SNR}
   \ENDFOR
  \RETURN Performance vectors: $Q(\mathbf{e}_c)$ (RMSE), $\bm{\xi}$[dB]
  \end{algorithmic} 
\end{algorithm}
% subsection results (end)

\begin{figure*}[!t]
\centerline{\subfloat[Butterworth case, $L=108$]{ \setlength\figureheight{1.9cm}
    \setlength\figurewidth{7cm}
    % This file was created by matlab2tikz v0.3.0.
% Copyright (c) 2008--2012, Nico Schlömer <nico.schloemer@gmail.com>
% All rights reserved.
% 
% The latest updates can be retrieved from
%   http://www.mathworks.com/matlabcentral/fileexchange/22022-matlab2tikz
% where you can also make suggestions and rate matlab2tikz.
% 
% 
% 
\pgfplotsset{compat=1.5}

\begin{tikzpicture}
\footnotesize
\begin{axis}[%
width=\figurewidth,
height=\figureheight,
tick label style={/pgf/number format/fixed},
scale only axis,
xmin=0, xmax=1.4,
ymin=0, ymax=1,
ylabel={Est. occurrence prob.},
yticklabel style={/pgf/number format/fixed, /pgf/number format/precision=3},
scale only axis,
name=plot1,
legend style={draw=black,fill=white,align=left},
area legend]
\addplot[ybar,bar width=0.0205809271977096\figurewidth,bar shift=-0.001500404635988548\figurewidth,fill=green,draw=black] plot coordinates{ (0.0181831987681414,0.047) (0.0453436773178482,0.953) (0.072504155867555,0) (0.0996646344172618,0) (0.126825112966969,0) (0.153985591516675,0) (0.181146070066382,0) (0.208306548616089,0) (0.235467027165796,0) (0.262627505715502,0) (0.289787984265209,0) (0.316948462814916,0) (0.344108941364623,0) (0.37126941991433,0) (0.398429898464036,0) (0.425590377013743,0) (0.45275085556345,0) (0.479911334113157,0) (0.507071812662863,0) (0.53423229121257,0) (0.561392769762277,0) (0.588553248311984,0) (0.615713726861691,0) (0.642874205411398,0) (0.670034683961104,0) (0.697195162510811,0) (0.724355641060518,0) (0.751516119610225,0) (0.778676598159931,0) (0.805837076709638,0) (0.832997555259345,0) (0.860158033809052,0) (0.887318512358759,0) (0.914478990908465,0) (0.941639469458172,0) (0.968799948007879,0) (0.995960426557586,0) (1.02312090510729,0) (1.050281383657,0) (1.07744186220671,0) };
\addlegendentry{{Calibrated impulse response,}\\ {$M_{\text{q}}=189$, $\sigma=2.7\cdot10^{-6}$}};

\addplot [
color=black,
solid,
forget plot
]
coordinates{
 (0,0)(1.4,0) 
};
\addplot[ybar,bar width=0.0205809271977096\figurewidth,bar shift=-0.001500404635988548\figurewidth,fill=blue,draw=black] plot coordinates{ (0.0181831987681414,0) (0.0453436773178482,0) (0.072504155867555,0) (0.0996646344172618,0.148333333333333) (0.126825112966969,0.748666666666667) (0.153985591516675,0.103) (0.181146070066382,0) (0.208306548616089,0) (0.235467027165796,0) (0.262627505715502,0) (0.289787984265209,0) (0.316948462814916,0) (0.344108941364623,0) (0.37126941991433,0) (0.398429898464036,0) (0.425590377013743,0) (0.45275085556345,0) (0.479911334113157,0) (0.507071812662863,0) (0.53423229121257,0) (0.561392769762277,0) (0.588553248311984,0) (0.615713726861691,0) (0.642874205411398,0) (0.670034683961104,0) (0.697195162510811,0) (0.724355641060518,0) (0.751516119610225,0) (0.778676598159931,0) (0.805837076709638,0) (0.832997555259345,0) (0.860158033809052,0) (0.887318512358759,0) (0.914478990908465,0) (0.941639469458172,0) (0.968799948007879,0) (0.995960426557586,0) (1.02312090510729,0) (1.050281383657,0) (1.07744186220671,0) };

\addlegendentry{{Calibrated impulse response,}\\ {$M_{\text{q}}=105$, $\sigma=1.2\cdot10^{-5}$}};
\end{axis}

\begin{axis}[%
width=\figurewidth,
height=\figureheight,
scale only axis,
xmin=0, xmax=1.4,
xlabel={Root Mean Square Error (RMSE) $\cdot 10^3$},
ymin=0, ymax=0.08,
scaled y ticks = false,
yticklabel style={/pgf/number format/fixed, /pgf/number format/precision=3},
ylabel={Est. occurrence prob.},
scale only axis,
at=(plot1.below south west), anchor=above north west,
legend style={draw=black,fill=white,align=left},
area legend]
\addplot[ybar,bar width=0.0205809271977096\figurewidth,bar shift=-0.0015004635988548\figurewidth,fill=red,draw=black] plot coordinates{ (0.0181831987681414,0.0183333333333333) (0.0453436773178482,0.0336666666666667) (0.072504155867555,0.0573333333333333) (0.0996646344172618,0.047) (0.126825112966969,0.0716666666666667) (0.153985591516675,0.054) (0.181146070066382,0.0573333333333333) (0.208306548616089,0.0543333333333333) (0.235467027165796,0.0563333333333333) (0.262627505715502,0.056) (0.289787984265209,0.0423333333333333) (0.316948462814916,0.0433333333333333) (0.344108941364623,0.0326666666666667) (0.37126941991433,0.0353333333333333) (0.398429898464036,0.037) (0.425590377013743,0.0343333333333333) (0.45275085556345,0.0286666666666667) (0.479911334113157,0.0246666666666667) (0.507071812662863,0.0243333333333333) (0.53423229121257,0.023) (0.561392769762277,0.0236666666666667) (0.588553248311984,0.024) (0.615713726861691,0.0176666666666667) (0.642874205411398,0.0163333333333333) (0.670034683961104,0.0136666666666667) (0.697195162510811,0.0113333333333333) (0.724355641060518,0.0133333333333333) (0.751516119610225,0.00866666666666667) (0.778676598159931,0.008) (0.805837076709638,0.00933333333333333) (0.832997555259345,0.00766666666666667) (0.860158033809052,0.00533333333333333) (0.887318512358759,0.00333333333333333) (0.914478990908465,0.00266666666666667) (0.941639469458172,0.000666666666666667) (0.968799948007879,0.000666666666666667) (0.995960426557586,0.000333333333333333) (1.02312090510729,0.001) (1.050281383657,0.000333333333333333) (1.07744186220671,0.000333333333333333) };

\addlegendentry{{Non-calibrated impulse response}\\{$\sigma=2.1\cdot10^{-4}$}};

\addplot [
color=black,
solid,
forget plot
]
coordinates{
 (0,0)(1.4,0) 
};
\end{axis}
\end{tikzpicture}%
\label{fig:figures_calibration-3000_butter_all-hist}}
\hfil
\subfloat[Chebyshev case, $L=228$]{\setlength\figureheight{1.9cm}
    \setlength\figurewidth{7cm}
    % This file was created by matlab2tikz v0.3.0.
% Copyright (c) 2008--2012, Nico Schlömer <nico.schloemer@gmail.com>
% All rights reserved.
% 
% The latest updates can be retrieved from
%   http://www.mathworks.com/matlabcentral/fileexchange/22022-matlab2tikz
% where you can also make suggestions and rate matlab2tikz.
% 
% 
% 
\begin{tikzpicture}

\pgfplotsset{compat=1.5}
\footnotesize
\begin{axis}[%
width=\figurewidth,
height=\figureheight,
scale only axis,
xmin=0, xmax=1.4,
ymin=0, ymax=1,
ylabel={Est. occurrence prob.},
yticklabel style={/pgf/number format/fixed, /pgf/number format/precision=3},
scale only axis,
name=plot1,
legend style={draw=black,fill=white,align=left},
area legend]
\addplot[ybar,bar width=0.0220809271977096\figurewidth,bar shift=-0.00244404635988548\figurewidth,fill=green,draw=black] plot coordinates{ (0.0249850194087497,0.994666666666667) (0.0554046737356276,0.00533333333333333) (0.0858243280625055,0) (0.116243982389383,0) (0.146663636716261,0) (0.177083291043139,0) (0.207502945370017,0) (0.237922599696895,0) (0.268342254023773,0) (0.298761908350651,0) (0.329181562677529,0) (0.359601217004407,0) (0.390020871331285,0) (0.420440525658162,0) (0.45086017998504,0) (0.481279834311918,0) (0.511699488638796,0) (0.542119142965674,0) (0.572538797292552,0) (0.60295845161943,0) (0.633378105946308,0) (0.663797760273186,0) (0.694217414600064,0) (0.724637068926942,0) (0.755056723253819,0) (0.785476377580697,0) (0.815896031907575,0) (0.846315686234453,0) (0.876735340561331,0) (0.907154994888209,0) (0.937574649215087,0) (0.967994303541965,0) (0.998413957868843,0) (1.02883361219572,0) (1.0592532665226,0) (1.08967292084948,0) (1.12009257517635,0) (1.15051222950323,0) (1.18093188383011,0) (1.21135153815699,0) };

\addlegendentry{{Calibrated impulse response,}\\ {$M_{\text{q}}=273$, $\sigma=3.1\cdot10^{-6}$}};

\addplot [
color=black,
solid,
forget plot
]
coordinates{
 (0,0)(1.4,0) 
};
\addplot[ybar,bar width=0.0220809271977096\figurewidth,bar shift=-0.00244404635988548\figurewidth,fill=blue,draw=black] plot coordinates{ (0.0249850194087497,0.151666666666667) (0.0554046737356276,0.311666666666667) (0.0858243280625055,0.207666666666667) (0.116243982389383,0.141333333333333) (0.146663636716261,0.0973333333333333) (0.177083291043139,0.0526666666666667) (0.207502945370017,0.0246666666666667) (0.237922599696895,0.0103333333333333) (0.268342254023773,0.00166666666666667) (0.298761908350651,0.001) (0.329181562677529,0) (0.359601217004407,0) (0.390020871331285,0) (0.420440525658162,0) (0.45086017998504,0) (0.481279834311918,0) (0.511699488638796,0) (0.542119142965674,0) (0.572538797292552,0) (0.60295845161943,0) (0.633378105946308,0) (0.663797760273186,0) (0.694217414600064,0) (0.724637068926942,0) (0.755056723253819,0) (0.785476377580697,0) (0.815896031907575,0) (0.846315686234453,0) (0.876735340561331,0) (0.907154994888209,0) (0.937574649215087,0) (0.967994303541965,0) (0.998413957868843,0) (1.02883361219572,0) (1.0592532665226,0) (1.08967292084948,0) (1.12009257517635,0) (1.15051222950323,0) (1.18093188383011,0) (1.21135153815699,0) };

\addlegendentry{{Calibrated impulse response,}\\ {$M_{\text{q}}=189$, $\sigma=4.9\cdot10^{-5}$}};

\end{axis}

\begin{axis}[%
width=\figurewidth,
height=\figureheight,
scale only axis,
xmin=0, xmax=1.4,
xlabel={Root Mean Square Error (RMSE) $\cdot 10^3$},
ymin=0, ymax=0.08,
scaled y ticks = false,
yticklabel style={/pgf/number format/fixed, /pgf/number format/precision=3},
ylabel={Est. occurrence prob.},
scale only axis,
at=(plot1.below south west), anchor=above north west,
legend style={draw=black,fill=white,align=left},
area legend]
\addplot[ybar,bar width=0.0220809271977096\figurewidth,bar shift=-0.00244404635988548\figurewidth,fill=red,draw=black] plot coordinates{ (0.0249850194087497,0.026) (0.0554046737356276,0.0556666666666667) (0.0858243280625055,0.0623333333333333) (0.116243982389383,0.0696666666666667) (0.146663636716261,0.064) (0.177083291043139,0.0553333333333333) (0.207502945370017,0.054) (0.237922599696895,0.0453333333333333) (0.268342254023773,0.0516666666666667) (0.298761908350651,0.0486666666666667) (0.329181562677529,0.046) (0.359601217004407,0.0426666666666667) (0.390020871331285,0.0346666666666667) (0.420440525658162,0.035) (0.45086017998504,0.0356666666666667) (0.481279834311918,0.0276666666666667) (0.511699488638796,0.0326666666666667) (0.542119142965674,0.0283333333333333) (0.572538797292552,0.0236666666666667) (0.60295845161943,0.0203333333333333) (0.633378105946308,0.0183333333333333) (0.663797760273186,0.0193333333333333) (0.694217414600064,0.0166666666666667) (0.724637068926942,0.014) (0.755056723253819,0.012) (0.785476377580697,0.012) (0.815896031907575,0.0106666666666667) (0.846315686234453,0.00766666666666667) (0.876735340561331,0.00666666666666667) (0.907154994888209,0.005) (0.937574649215087,0.00766666666666667) (0.967994303541965,0.004) (0.998413957868843,0.00133333333333333) (1.02883361219572,0.00266666666666667) (1.0592532665226,0.000666666666666667) (1.08967292084948,0.000333333333333333) (1.12009257517635,0) (1.15051222950323,0.001) (1.18093188383011,0) (1.21135153815699,0.000666666666666667) };

\addlegendentry{{Non-calibrated impulse response}\\{$\sigma=2.3\cdot10^{-4}$}};

\addplot [
color=black,
solid,
forget plot
]
coordinates{
 (0,0)(1.4,0) 
};
\end{axis}
\end{tikzpicture}%
\label{fig:chebysh_calibration-3000-LSCOV-hist}}}
\caption{Calibration performance on the random demodulator using a filter with components subject to deviations of 2 \% for capacitors and inductors. Calibration methods \eqref{eq:least-squares} and \eqref{eq:least-squares-TK} were conducted for cases $M_{\text{q}} > L$ and $M_{\text{q}} < L$ respectively.}
\label{MC_histograms}
\end{figure*}
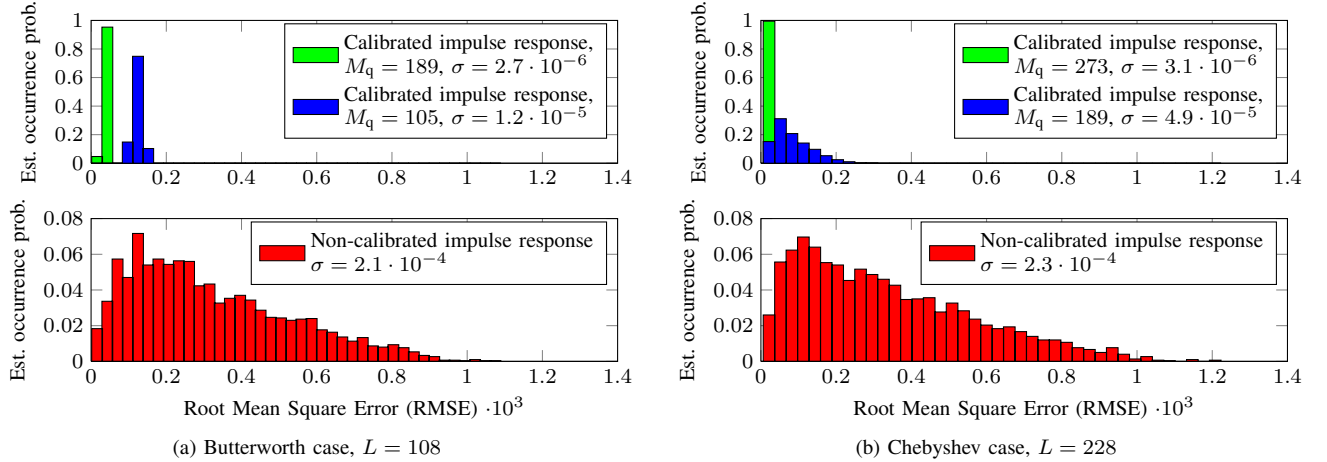

The mean value of $Q({\rm \mathbf{e}_{\rm p}})$ over the entire simulation for the Butterworth approximation was computed to $3.22\cdot 10^{-4}$. After calibration \eqref{eq:least-squares} using $M = 189$ samples, the mean value of $Q({\rm {\mathbf{\hat{e}}_{\rm p}}})$ equals $3.6\cdot 10^{-5}$, which is $\approx 9$ times lower than the mean RMSE of the initial perturbation $Q({\rm \mathbf{e}_{\rm p}})$. The calibration expressed by \eqref{eq:least-squares-TK} performed by taking $M = 105$ compressed samples results in mean RMSE of $1.18\cdot 10^{-4}$, which on average is 2.7 times smaller than the mean error of the initial perturbations. 
The impulse response of the Chebyshev approximated filter was represented by $L=228$ samples. The mean RMSE of the simulated perturbation resulted in $3.41\cdot 10^{-4}$. Utilizing Algorithm~\ref{alg1} with $M = 273$ samples results in mean $Q({\rm \mathbf{e}_{\rm p}}) = 3.23\cdot 10^{-5}$ which is $\approx 9$ times smaller. Calibration performed by taking $M < L$ (Fig.~\ref{MC_histograms}\subref{fig:chebysh_calibration-3000-LSCOV-hist}) reduces the RMSE to $8.77\cdot 10^{-5}$, which on average is 3.9 times smaller than the mean error of the initial perturbations.

We also evaluated the reconstruction quality under different error sizes using  Algorithms \ref{alg2} and \ref{alg3}. The results of the reconstruction with and without calibration are presented in Table~\ref{tab:reference_perfoance}. The table columns show the minimum (1), average (2), and maximum (3) recorded error, respectively, within 3000 simulated cases and corresponding SNR values for each error.

\begin{table}[h!]
\centering
\renewcommand{\arraystretch}{1.3}
\caption{Impulse response RMSE and corresponding reconstructed signal SNR (Chebyshev filter).}
\begin{tabular}{|l|c|c|c|c|c|c|}
\hline
			&\multicolumn{3}{c|}{non-calibrated}&\multicolumn{3}{c|}{calibrated with $M_{\text{q}}=273$}\\
			 \hline
			& \multicolumn{3}{c|}{case number} & \multicolumn{3}{c|}{case number}\\
			 \hline
	 	& 1 & 2& 3 & 1 & 2 & 3 \\ 
			\hline
RMSE $\cdot {10^{4}}$ 		& \num[round-mode = places,add-decimal-zero = true, add-integer-zero = true, round-precision = 2]{0.096515}    & \num[round-mode = places,add-decimal-zero = true, add-integer-zero = true, round-precision = 2]{2.931424}	  & \num[round-mode = places,add-decimal-zero = true, add-integer-zero = true, round-precision = 2]{12.113515} 	 & ~\num[round-mode = places,add-decimal-zero = true, add-integer-zero = true, round-precision = 2]{0.2161}~ 		& ~\num[round-mode = places,add-decimal-zero = true, add-integer-zero = true, round-precision = 2]{0.2555}~ 		& \num[round-mode = places,add-decimal-zero = true, add-integer-zero = true, round-precision = 2]{0.1852} \\
\hline
SNR \si{\deci\bel}    &\num[round-mode = places,add-decimal-zero = true, add-integer-zero = true, round-precision = 1]{68.0160}  			&\num[round-mode = places,add-decimal-zero = true, add-integer-zero = true, round-precision = 1]{38.2941}  		  	  &\num[round-mode = places,add-decimal-zero = true, add-integer-zero = true, round-precision = 1]{25.8667} 			&\num[round-mode = places,add-decimal-zero = true, add-integer-zero = true, round-precision = 1]{75.9449} 	    	  &\num[round-mode = places,add-decimal-zero = true, add-integer-zero = true, round-precision = 1]{75.0853} 			&\num[round-mode = places,add-decimal-zero = true, add-integer-zero = true, round-precision = 1]{78.1725} \\
\hline
\end{tabular}
\label{tab:reference_perfoance}
\end{table}

% \vspace{-0.5cm}
We have investigated the performance of \eqref{eq:least-squares} with respect to the amount of samples used in the least-squares estimation. Using the Butterworth approximated filter architecture, we modeled the impulse response with $L=108$ and performed 11 calibration schemes \eqref{eq:least-squares} with different $M_{\rm q} \in \{42, 63, 105, 126, 189, 315, 630, 1050, 2100, 4200, 8400 \}$. The results are shown in Figure~\ref{fig:calibSize}. The tests utilized calibrating signals with $K\in \{5, 10, 50\}$ tones.
Table~\ref{tab:rmse_calSize} juxtaposes the calibration performance in terms of computation time and RMSE. The initial error size in the impulse response was $56.57\cdot 10^{-5}$.

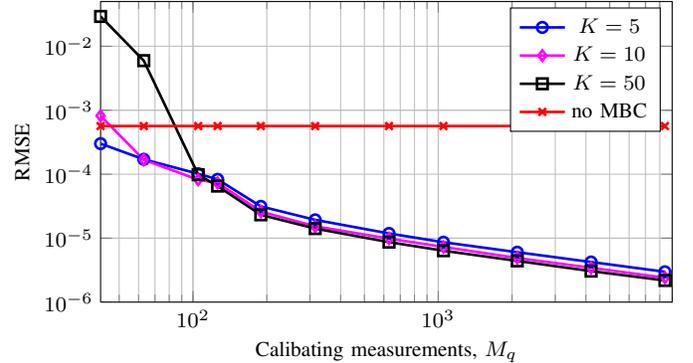
\begin{figure}[h!]  
    \setlength\figureheight{4cm}
    \setlength\figurewidth{7.6cm}
 % This file was created by matlab2tikz v0.3.0.
% Copyright (c) 2008--2012, Nico Schlömer <nico.schloemer@gmail.com>
% All rights reserved.
% 
% The latest updates can be retrieved from
%   http://www.mathworks.com/matlabcentral/fileexchange/22022-matlab2tikz
% where you can also make suggestions and rate matlab2tikz.
% 
% 
% 

% defining custom colors
\definecolor{mycolor1}{rgb}{1,0,1}

\begin{tikzpicture}
\footnotesize

\begin{semilogyaxis}[%
 /pgf/number format/.cd,
	set thousands separator={},
	set decimal separator={.},
width=\figurewidth,
height=\figureheight,
scale only axis,
ymode=log,
xmode=log,
xmin=42, xmax=9000,
xlabel={Calibating measurements, $M_q$},
ymin=1e-06, ymax=0.05,
yminorticks=true,
ylabel={RMSE},
grid=both,
ticks=both,
minor x tick num={1},
minor ytick={0.000001,0.00001, 0.001},
legend style={draw=black,fill=white,align=left}]
\addplot [
color=blue,
solid,
mark=o,
line width=1pt,
mark options={solid}
]
coordinates{
 (42,0.000300223956451391)(63,0.000170773768806204)(105,0.000101517745294379)(126,8.29782771924623e-05)(189,3.14109294566264e-05)(315,1.93282455613497e-05)(630,1.18567647506037e-05)(1050,8.64326917925979e-06)(2100,6.03521886618472e-06)(4200,4.24081401208919e-06)(8400,2.98446896137967e-06) 
};
\addlegendentry{$K=5$};

\addplot [
color=mycolor1,
solid,
mark=diamond,
line width=1pt,
mark options={solid}
]
coordinates{
 (42,0.000817336914892643)(63,0.000167986701823228)(105,8.19217852432629e-05)(126,7.29883248730082e-05)(189,2.60793032545632e-05)(315,1.53475897437479e-05)(630,9.84626752524098e-06)(1050,7.2882506013494e-06)(2100,4.9078626764389e-06)(4200,3.44313540317236e-06)(8400,2.41153015021672e-06) 
};
\addlegendentry{$K=10$};

\addplot [
color=black,
solid,
mark=square,
line width=1pt,
mark options={solid}
]
coordinates{
 (42,0.0291900469758571)(63,0.00595217463855818)(105,9.82995093512103e-05)(126,6.52146010137267e-05)(189,2.31881601389069e-05)(315,1.41356051983619e-05)(630,8.6794282406707e-06)(1050,6.3737275263669e-06)(2100,4.41351638543654e-06)(4200,3.05578625103689e-06)(8400,2.18259080035444e-06) 
};
\addlegendentry{$K=50$};

\addplot [
color=red,
solid,
mark=x,
line width=1pt,
mark options={solid}
]
coordinates{
 (42,0.000565742610649445)(63,0.000565742610649445)(105,0.000565742610649445)(126,0.000565742610649445)(189,0.000565742610649445)(315,0.000565742610649445)(630,0.000565742610649445)(1050,0.000565742610649445)(2100,0.000565742610649445)(4200,0.000565742610649445)(8400,0.000565742610649445) 
};
\addlegendentry{no MBC};

\end{semilogyaxis}
\end{tikzpicture}%	
	\caption{Calibration performance of the impulse response in RMSE versus the amount of samples used in the least-square estimation \eqref{eq:least-squares}}	
		\label{fig:calibSize}
\end{figure}

\vspace{-0.2cm}
\begin{table}[htb]
\centering
\renewcommand{\arraystretch}{1.3}
\caption{RMSE of the calibrated impulse response and the actual one.}
\begin{tabular}{|@{\hspace{0.1cm}}l@{\hspace{0.1cm}}|c|c|c|c|c|c|c|}
			\hline
& \multirow{2}{*}{K} 	& \multicolumn{6}{c|}{$M_{\rm q}$ samples}\\
			\cline{3-8}
	 	&  & 42 & 63& 105 & 126 & 1050 & 8400 \\ 
			\hline
  \multirow{3}{*}{RMSE $\cdot {10^{5}}$} & 5 &\num[round-mode = places,add-decimal-zero = true, add-integer-zero = true, round-precision = 2]{30.022} & 		\num[round-mode = places,add-decimal-zero = true, add-integer-zero = true, round-precision = 2]{17.077}&		 \num[round-mode = places,add-decimal-zero = true, add-integer-zero = true, round-precision = 2]{10.152} & 		  \num[round-mode = places,add-decimal-zero = true, add-integer-zero = true, round-precision = 2]{8.298} &		   \num[round-mode = places,add-decimal-zero = true, add-integer-zero = true, round-precision = 2]{1.933} & 	     \num[round-mode = places,add-decimal-zero = true, add-integer-zero = true, round-precision = 2]{0.298} \\
\cline{2-8}
& 10 & 						\num[round-mode = places,add-decimal-zero = true, add-integer-zero = true, round-precision = 2]{81.734} & 		\num[round-mode = places,add-decimal-zero = true, add-integer-zero = true, round-precision = 2]{16.799}&		 \num[round-mode = places,add-decimal-zero = true, add-integer-zero = true, round-precision = 2]{8.1922} & 		  \num[round-mode = places,add-decimal-zero = true, add-integer-zero = true, round-precision = 2]{7.2988} &		   \num[round-mode = places,add-decimal-zero = true, add-integer-zero = true, round-precision = 2]{0.72883} & 	     \num[round-mode = places,add-decimal-zero = true, add-integer-zero = true, round-precision = 2]{0.24115} \\
\cline{2-8}
& 50 & 						\num[round-mode = places,add-decimal-zero = true, add-integer-zero = true, round-precision = 2]{2919.005} & 		\num[round-mode = places,add-decimal-zero = true, add-integer-zero = true, round-precision = 2]{595.217}&		 \num[round-mode = places,add-decimal-zero = true, add-integer-zero = true, round-precision = 2]{9.830} & 		  \num[round-mode = places,add-decimal-zero = true, add-integer-zero = true, round-precision = 2]{6.521} &		   \num[round-mode = places,add-decimal-zero = true, add-integer-zero = true, round-precision = 2]{1.414} & 	     \num[round-mode = places,add-decimal-zero = true, add-integer-zero = true, round-precision = 2]{0.218} \\
\hline
  \multirow{3}{*}{time [\si{\second}]} &	5 & 													\num[round-mode = places,add-decimal-zero = true, add-integer-zero = true, round-precision = 2]{0.4509}  & \num[round-mode = places,add-decimal-zero = true, add-integer-zero = true, round-precision = 2]{0.4537}  & \num[round-mode = places,add-decimal-zero = true, add-integer-zero = true, round-precision = 2]{0.5054} &  \num[round-mode = places,add-decimal-zero = true, add-integer-zero = true, round-precision = 2]{0.0688} &  \num[round-mode = places,add-decimal-zero = true, add-integer-zero = true, round-precision = 2]{0.0746}  & \num[round-mode = places,add-decimal-zero = true, add-integer-zero = true, round-precision = 2]{3.5649}  \\
  \cline{2-8}
  &	10 &						  \num[round-mode = places,add-decimal-zero = true, add-integer-zero = true, round-precision = 2]{0.4520}  & \num[round-mode = places,add-decimal-zero = true, add-integer-zero = true, round-precision = 2]{0.4533}  & \num[round-mode = places,add-decimal-zero = true, add-integer-zero = true, round-precision = 2]{0.4922} &  \num[round-mode = places,add-decimal-zero = true, add-integer-zero = true, round-precision = 2]{0.0686} &  \num[round-mode = places,add-decimal-zero = true, add-integer-zero = true, round-precision = 2]{0.1343}  & \num[round-mode = places,add-decimal-zero = true, add-integer-zero = true, round-precision = 2]{3.5598}  \\
  \cline{2-8}
    &			50 			 &  \num[round-mode = places,add-decimal-zero = true, add-integer-zero = true, round-precision = 2]{0.4615}  & \num[round-mode = places,add-decimal-zero = true, add-integer-zero = true, round-precision = 2]{0.4541}  & \num[round-mode = places,add-decimal-zero = true, add-integer-zero = true, round-precision = 2]{0.4862} &  \num[round-mode = places,add-decimal-zero = true, add-integer-zero = true, round-precision = 2]{0.0685} &  \num[round-mode = places,add-decimal-zero = true, add-integer-zero = true, round-precision = 2]{0.0742}  & \num[round-mode = places,add-decimal-zero = true, add-integer-zero = true, round-precision = 2]{3.5602}  \\
\hline
\end{tabular}
\label{tab:rmse_calSize}
\end{table}

\begin{figure*}[t]
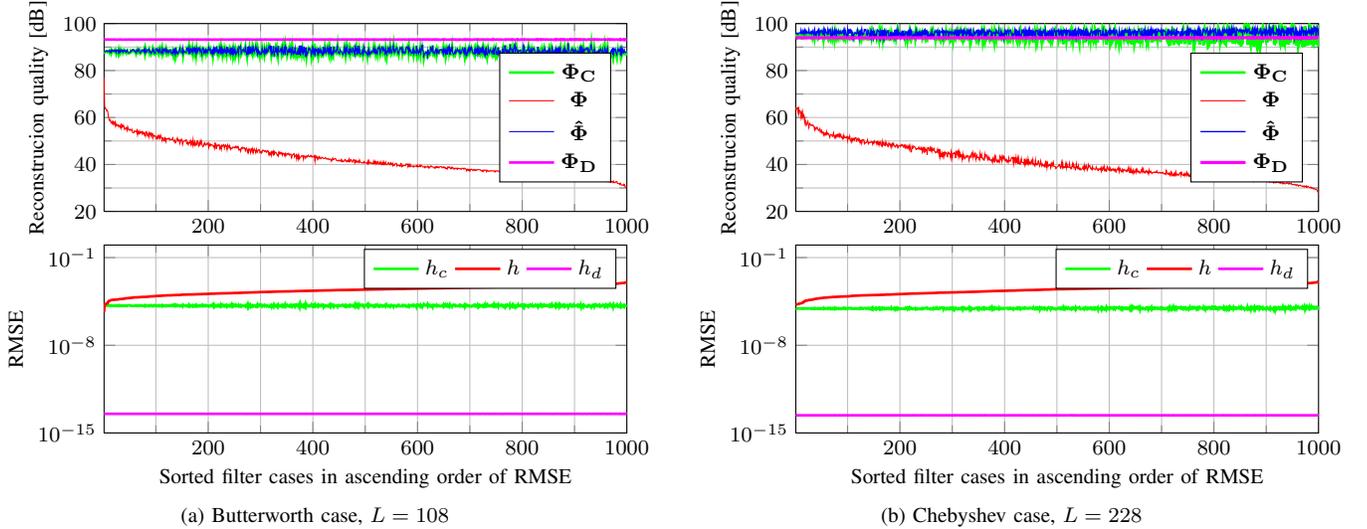

\centerline{\subfloat[Butterworth case, $L=108$]{ \setlength\figureheight{2.5cm}
    \setlength\figurewidth{6.95cm}
 \input{benchAnalysis_rdCalib_1x_fs12K_butter_SCOUT_cRMSE_1K.tikz}	
	\label{fig:benchAnalysis_butterScout12K_TIKZ}}
\hfil
\subfloat[Chebyshev case, $L=228$]{\setlength\figureheight{2.5cm}
    \setlength\figurewidth{6.95cm}
 \input{benchAnalysis_rdCalib_1x_fs12K_chebysh_SCOUT_cRMSE_1K.tikz}	
\label{fig:benchAnalysis_chebyshScout12K_TIKZ}}}
\caption{MBC vs. DFTTI benchmark: $f_N =$ 12~\si{\kilo\hertz}, $M_{\rm q}=1050$. Four SPGL1 reconstructions were performed with different measurement matrices:  $\bPhi_{\text{C}}$ - calibrated, $\bPhi$ - ideally modeled, $\hat{\bPhi}$ - perturbed (oracle) and $\bPhi_{\text{D}}$ - DFTTI obtained. RMSE values were calculated between the perturbed (oracle) impulse response and: calibrated $h_c$, ideally modeled $h$, and DFTTI obtained $h_d$. }
\label{benchAnalysis}
\end{figure*}

\vspace{-0.2cm}
\subsection{Benchmarking} % (fold)
\label{sub:benchmarking}
Finally, we benchmarked the MBC against the DFTTI method proposed in~\cite{becker2011}. The simulation set-up consisted of the same set of initial parameters as used in previous Monte Carlo simulations. Essentially Algorithms \ref{alg2} and \ref{alg3} were utilized to assess the reconstruction quality of each method.  
Filter realizations (Butterworth and Chebyshev) were subject to component nominal value variation according to \eqref{eq:truncg}. We generated 1000 deviating sets of components and sorted the data according to the resulting RMSE.  
We have recorded SNR, RMSE and the time taken to generate the calibrated measurement matrix $\bm{{\Phi_\text{C}}}$ with model-based calibration or $\bm{\Phi_\text{D}}$ using DFTTI. Table~\ref{tab:main_Bench} juxtaposes the results of the benchmark and 
 Fig.~\ref{benchAnalysis} visualizes them.

\begin{table}[htb!]
\centering
\renewcommand{\arraystretch}{1.3}
\caption{RMSE, SNR of the calibrated impulse response and time taken by the procedures: model-based calibration (MBC) and DFTTI.}
\begin{tabular}{|@{\hspace{0.1cm}}l@{\hspace{0.1cm}}|c|c|c|c|c|c|}
\hline
			&\multicolumn{2}{c|}{RMSE $\cdot 10^7$}&\multicolumn{2}{c|}{SNR [dB]}&\multicolumn{2}{c|}{time [s]}\\
			\hline
	 Method	& DFFTI & MBC& DFTTI & MBC & DFTTI & MBC \\ 
			\hline
Butter. & \num[round-mode = places,add-decimal-zero = true, add-integer-zero = true, round-precision = 2]{3.3824}$\cdot 10^{-7}$ & \num[round-mode = places,add-decimal-zero = true, add-integer-zero = true, round-precision = 2]{144.73}& \num[round-mode = places,add-decimal-zero = true, add-integer-zero = true, round-precision = 1]{93.1777} & \num[round-mode = places,add-decimal-zero = true, add-integer-zero = true, round-precision = 1]{88.0271} & \num[round-mode = places,add-decimal-zero = true, add-integer-zero = true, round-precision = 2]{173.7526} & \num[round-mode = places,add-decimal-zero = true, add-integer-zero = true, round-precision = 2]{0.1312} \\
\hline
Chebysh.  &  \num[round-mode = places,add-decimal-zero = true, add-integer-zero = true, round-precision = 2]{2.5677}$\cdot 10^{-7}$& \num[round-mode = places,add-decimal-zero = true, add-integer-zero = true, round-precision = 2]{88.91}  & \num[round-mode = places,add-decimal-zero = true, add-integer-zero = true, round-precision = 1]{93.9474} &  \num[round-mode = places,add-decimal-zero = true, add-integer-zero = true, round-precision = 1]{94.6646} &  \num[round-mode = places,add-decimal-zero = true, add-integer-zero = true, round-precision = 2]{180.9757}  & \num[round-mode = places,add-decimal-zero = true, add-integer-zero = true, round-precision = 2]{0.1457}  \\
\hline
\end{tabular}
\label{tab:main_Bench}
\end{table}

The convex solver SPGL1 used to execute the simulations was limited to perform
maximum 2500 iterations. This was done to impose a fair reconstruction time limit. The value is sufficiently high to allow perfect reconstruction within the limit, when we solve well-conditioned CS problems ("nice" $\bPhi$ and $\bPhi\bPsi$ RIP fulfillment)\cite{eldar2012compressed, Candes:2008uc}.

% subsection benchmarking (end)
\subsection{Discussion} % (fold)
\label{sub:discussion}
The presented results confirm that the calibration method compensates for filter modeling discrepancies. The method is most reliable in cases of taking a higher amount of samples than the impulse response is represented with ($M_{\rm q}> L$).
We have not observed any problems with the stability of the calibration formulation in \eqref{eq:least-squares}. On the contrary, when using \eqref{eq:least-squares-TK} for $M_{\rm q} \leq L$, we recorded  cases where the RMSE of the calibrated impulse response $Q({\rm {\mathbf{\hat e}}_{\rm p}}) \geq Q({\rm \mathbf{e}_{\rm p}})$. The  success rate of the approach downgrades rapidly with decreasing amount of samples and should be considered only when it is infeasible to gather $M_{\rm q} \geq L$. Also, when taking low amounts of samples, both methods are unable to correct the smallest errors. 
The data shows that already for $M_{\rm q} \approx 1.2\cdot L$, \eqref{eq:least-squares}  significantly decreases the error in the impulse response, contributing further to the reconstruction quality improvement. 
Furthermore, the method performance can be tuned by increasing the amount of calibrating signal $\x_{\rm q}$ tones $K$, as can be seen in Fig.~\ref{fig:calibSize}. This is related to the condition number of the matrix $\mathbf{D}$ in \eqref{eq:D-matrix} and modeling density of $\x$ and $\mathbf{p}$. However, the method in \eqref{eq:least-squares-TK} showed performance degradation for increasing $K$.
The method~\eqref{eq:least-squares} enables sufficient correction of the impulse response but it does not reach the same precision as the DFTTI method. Also, because of the fact that the framework operates on the truncated impulse responses, the reconstruction is more susceptible to component imperfections, which can be observed in Fig.~\ref{benchAnalysis}. The SNR of the DFTTI method is very stable regardless of the impulse response error. 
 It is important to notice though, that the proposed method requires only an order of $M$ samples to carry out successful calibration as opposed to $M\times N$ used by DFTTI.  For the problem of size $800 \times 12600$, DFTTI took $~12.6\,\si{\kilo}$ samples more than the proposed method, which was the main time-limiting factor. Both of the methods used the same RD signal acquisition framework to facilitate fair time comparison.  This makes the model-based calibration method very suitable for systems that require frequent re-calibration.

\section{Conclusion} % (fold)
\label{sec:conclusion}
In this article, we presented a supervised model-based calibration method for the random demodulator framework. The calibration addresses the measurement matrix discrepancy that appears when an unaccounted change of the filter characteristics occurs in the analog front-end of the random demodulator architecture. With the assumption of a known initial filter model, the method exploits the nature of the error and identifies it through linear estimation. The amount of samples necessary to assure successful calibration was orders of magnitude lower when compared to the existing techniques. Through a series of numerical experiments  we have shown that the method works independently of the filter realization, and can be used universally as a calibration step before commencing acquisition and reconstruction. The calibration was observed to minimize the error to a level that it was insignificant in affecting the reconstruction quality. This increased the reconstruction of noiseless signal up to 50~\si{\decibel}.
The method does not require any modifications to the hardware or its operational frequency, making it easy to implement.  
% section conclusion (end)

\section*{Acknowledgment}
The authors would like to thank S. Becker for sharing the DFTTI calibration code utilized in the RMPI framework\cite{becker2011}.

\ifCLASSOPTIONcaptionsoff 
\newpage \fi

\bibliographystyle{IEEEtran}
\bibliography{bibliography}

\begin{thebibliography}{10}
\providecommand{\url}[1]{#1}
\csname url@samestyle\endcsname
\providecommand{\newblock}{\relax}
\providecommand{\bibinfo}[2]{#2}
\providecommand{\BIBentrySTDinterwordspacing}{\spaceskip=0pt\relax}
\providecommand{\BIBentryALTinterwordstretchfactor}{4}
\providecommand{\BIBentryALTinterwordspacing}{\spaceskip=\fontdimen2\font plus
\BIBentryALTinterwordstretchfactor\fontdimen3\font minus
  \fontdimen4\font\relax}
\providecommand{\BIBforeignlanguage}[2]{{%
\expandafter\ifx\csname l@#1\endcsname\relax
\typeout{** WARNING: IEEEtran.bst: No hyphenation pattern has been}%
\typeout{** loaded for the language `#1'. Using the pattern for}%
\typeout{** the default language instead.}%
\else
\language=\csname l@#1\endcsname
\fi
#2}}
\providecommand{\BIBdecl}{\relax}
\BIBdecl

\bibitem{Candes2006b}
E.~J. Cand{\`e}s, ``{Compressive sampling},'' \emph{Proceedings of the
  International Congress of Mathematicians: Madrid, August 22-30, 2006: invited
  lectures}, pp. 1433--1452, 2006.

\bibitem{Donoho:2006vb}
D.~Donoho, ``{Compressed Sensing},'' \emph{IEEE Transactions on Information
  Theory}, vol.~52, no.~4, pp. 1289--1306, 2006.

\bibitem{Candes:2008uc}
E.~J. Cand{\`e}s and M.~B. Wakin, ``{An Introduction To Compressive
  Sampling},'' \emph{IEEE Signal Processing Magazine}, vol.~25, no.~2, pp.
  21--30, 2008.

\bibitem{Kirolos:2006p3806}
S.~Kirolos, J.~Laska, M.~Wakin, M.~Duarte, D.~Baron, T.~Ragheb, Y.~Massoud, and
  R.~Baraniuk, ``{Analog-to-information conversion via random demodulation},''
  \emph{Workshop on Design, Applications, Integration and Software, IEEE
  Dallas/CAS}, pp. 71--74, 2006.

\bibitem{Laska:2007p3817}
J.~Laska, S.~Kirolos, M.~F. Duarte, T.~Ragheb, R.~G. Baraniuk, and Y.~Massoud,
  ``{Theory and implementation of an analog-to-information converter using
  random demodulation},'' \emph{IEEE International Symposium on Circuits and
  Systems, ISCAS.}, pp. 1959--1962, 2007.

\bibitem{Mishali:2010p3825}
M.~Mishali and Y.~C. Eldar, ``{From theory to practice: Sub-Nyquist sampling of
  sparse wideband analog signals},'' \emph{IEEE Journal of Selected Topics in
  Signal Processing}, vol.~4, no.~2, pp. 375--391, 2010.

\bibitem{Mishali:2010p4876}
M.~Mishali and Y.~Eldar, ``Xampling: Analog data compression,'' in \emph{Data
  Compression Conference (DCC)}, 2010, pp. 366--375.

\bibitem{Ragheb:2008vb}
T.~Ragheb, J.~Laska, H.~Nejati, S.~Kirolos, R.~G. Baraniuk, and Y.~Massoud,
  ``{A prototype hardware for random demodulation based compressive
  analog-to-digital conversion},'' \emph{51st Midwest Symposium on Circuits and
  Systems, MWSCAS.}, pp. 37--40, 2008.

\bibitem{Yang:2009vb}
D.~Yang, H.~Li, G.~Peterson, and A.~Fathy, ``{Compressed sensing based UWB
  receiver: Hardware compressing and FPGA reconstruction},'' in \emph{43rd
  Annual Conference on Information Sciences and Systems, CISS.}, 2009, pp.
  198--201.

\bibitem{becker2011}
\BIBentryALTinterwordspacing
S.~R. Becker, ``Practical compressed sensing : modern data acquisition and
  signal processing,'' Thesis (Dissertation (Ph.D.)), California Institute of
  Technology, Jun. 2011. [Online]. Available:
  \url{http://resolver.caltech.edu/CaltechTHESIS:06022011-152525054}
\BIBentrySTDinterwordspacing

\bibitem{Unser:2000vo}
M.~Unser, ``{Sampling-50 years after Shannon},'' \emph{Proceedings of the
  IEEE}, vol.~88, no.~4, pp. 569--587, 2000.

\bibitem{Shannon:1949:CPN}
C.~Shannon, ``Communication in the presence of noise,'' \emph{Proceedings of
  the IRE}, vol.~37, no.~1, pp. 10--21, jan. 1949.

\bibitem{tropp2010c}
J.~Tropp and S.~Wright, ``Computational methods for sparse solution of linear
  inverse problems,'' \emph{Proceedings of the IEEE}, vol.~98, no.~6, pp.
  948--958, Jun. 2010.

\bibitem{Tropp:2010p3813}
J.~A. Tropp, J.~N. Laska, M.~F. Duarte, J.~K. Romberg, and R.~G. Baraniuk,
  ``{Beyond Nyquist: Efficient Sampling of Sparse Bandlimited Signals},''
  \emph{IEEE Transactions on Information Theory}, vol.~56, no.~1, pp. 520--544,
  2010.

\bibitem{Anonymous:ctmnJhsb}
T.~P. Boufounos and M.~S. Asif, ``{Compressive Sensing for streaming signals
  using the Streaming Greedy Pursuit},'' 2010, pp. 1205--1210.

\bibitem{eldar2012compressed}
Y.~C. Eldar and G.~Kutyniok, \emph{{Compressed Sensing: Theory and
  Applications}}.\hskip 1em plus 0.5em minus 0.4em\relax Cambridge University
  Press, 2012.

\bibitem{pankiewicz11}
P.~J. Pankiewicz, T.~Arildsen, and T.~Larsen, ``{Sensitivity of the Random
  Demodulation Framework to Filter Tolerances},'' in \emph{Proceedings of the
  European Signal Processing Conference (EUSIPCO)}, 2011, pp. 534--538.

\bibitem{Gribonval12}
R.~Gribonval, G.~Chardon, and L.~Daudet, ``Blind calibration for compressed
  sensing by convex optimization,'' in \emph{IEEE International Conference on
  Acoustics, Speech and Signal Processing (ICASSP)}, march 2012, pp.
  2713--2716.

\bibitem{Herman2010}
M.~A. Herman and T.~Strohmer, ``{General Deviants: An Analysis of Perturbations
  in Compressed Sensing},'' \emph{IEEE Journal of Selected Topics in Signal
  Processing}, vol.~4, no.~2, pp. 342--349, 2010.

\bibitem{Herman:2010td}
M.~Herman and D.~Needell, ``Mixed operators in compressed sensing,'' in
  \emph{44th Annual Conference on Information Sciences and Systems, CISS}, Mar.
  2010, pp. 1--6.

\bibitem{Wang:2011kc}
Q.~Wang and Z.~Liu, ``{Sampling Matrix Perturbation Analysis of Subspace
  Pursuit for Compressive Sensing},'' in \emph{Information and
  Automation}.\hskip 1em plus 0.5em minus 0.4em\relax Berlin, Heidelberg:
  Springer Berlin Heidelberg, 2011, pp. 581--588.

\bibitem{Rosenbaum2010}
M.~Rosenbaum and A.~B. Tsybakov, ``{Sparse recovery under matrix
  uncertainty},'' \emph{The Annals of Statistics}, vol.~38, no.~5, pp.
  2620--2651, Oct. 2010.

\bibitem{Zhu2011}
H.~Zhu, G.~Leus, and G.~B. Giannakis, ``{Sparsity-Cognizant Total Least-Squares
  for Perturbed Compressive Sampling},'' \emph{IEEE Transactions on Signal
  Processing}, vol.~59, no.~5, pp. 2002--2016, 2011.

\bibitem{Liu:2010tc}
Y.~Liu and Q.~Wan, ``Anti-measurement matrix uncertainty for robust sparse
  signal recovery with the mixed l2 and l1 norms constraint,'' \emph{ArXiv
  pre-print}, vol. abs/1006.0054, 2010.

\bibitem{Han:2011eh}
X.~Han, H.~Zhang, and H.~Meng, ``{TLS-FOCUSS for sparse recovery with perturbed
  dictionary},'' in \emph{IEEE International Conference on Acoustics, Speech
  and Signal Processing (ICASSP)}, 2011, pp. 3952--3955.

\bibitem{Davies:2008dk}
M.~E. Davies and T.~Blumensath, ``{Faster {\&} greedier: algorithms for sparse
  reconstruction of large datasets},'' \emph{3rd International Symposium on
  Communications, Control and Signal Processing, ISCCSP.}, pp. 774--779, 2008.

\bibitem{tropp2007}
J.~A. Tropp and A.~Gilbert, ``{Signal Recovery From Random Measurements Via
  Orthogonal Matching Pursuit},'' \emph{IEEE Transactions on Information
  Theory}, vol.~53, no.~12, pp. 4655--4666, 2007.

\bibitem{Dai2009b}
W.~Dai and O.~Milenkovic, ``{Subspace Pursuit for Compressive Sensing Signal
  Reconstruction},'' \emph{IEEE Transactions on Information Theory}, vol.~55,
  no.~5, pp. 2230--2249, 2009.

\bibitem{Chen2001}
S.~S. Chen, D.~L. Donoho, and M.~A. Saunders, ``Atomic decomposition by basis
  pursuit,'' \emph{SIAM Journal on Scientific Computing}, vol.~20, pp. 33--61,
  1998.

\bibitem{huelsman1993active}
L.~Huelsman, \emph{Active and Passive Analog Filter Design: An Introduction},
  ser. McGraw-Hill series in electrical and computer engineering: Electronics
  and VLSI circuits.\hskip 1em plus 0.5em minus 0.4em\relax McGraw-Hill, Inc.,
  1993.

\bibitem{oppenheim2010discrete}
A.~V. Oppenheim, R.~W. Schafer, and J.~R. Buck, \emph{Discrete-time signal
  processing (2nd ed.)}.\hskip 1em plus 0.5em minus 0.4em\relax Upper Saddle
  River, NJ, USA: Prentice-Hall, Inc., 1999.

\bibitem{kinget1996}
P.~Kinget and M.~Steyaert, ``Impact of transistor mismatch on the
  speed-accuracy-power trade-off of analog cmos circuits,'' in
  \emph{Proceedings of the IEEE Custom Integrated Circuits Conference}, May
  1996, pp. 333--336.

\bibitem{steyaert1997}
M.~Steyaert, V.~Peluso, J.~Bastos, P.~Kinget, and W.~Sansen, ``Custom analog
  low power design: the problem of low voltage and mismatch,'' in
  \emph{Proceedings of the IEEE Custom Integrated Circuits Conference}, May
  1997, pp. 285--292.

\bibitem{kinget2005}
P.~Kinget, ``Device mismatch and tradeoffs in the design of analog circuits,''
  \emph{IEEE Journal of Solid-State Circuits}, vol.~40, no.~6, pp. 1212--1224,
  Jun. 2005.

\bibitem{Filanovsky:2012be}
I.~Filanovsky, ``{Sensitivity and Selectivity},'' in \emph{Circuits {\&}
  Filters Handbook 3e}.\hskip 1em plus 0.5em minus 0.4em\relax Boca Raton,
  Florida: CRC Press, LLC, Jan. 2012.

\bibitem{Candes2006c}
E.~J. Cand{\`e}s, J.~K. Romberg, and T.~Tao, ``{Stable signal recovery from
  incomplete and inaccurate measurements},'' \emph{Communications on Pure and
  Applied Mathematics}, vol.~59, no.~8, pp. 1207--1223, 2006.

\bibitem{Donoho:2006ja}
D.~Donoho, M.~Elad, and V.~Temlyakov, ``{Stable recovery of sparse overcomplete
  representations in the presence of noise},'' \emph{IEEE Transactions
  onInformation Theory}, vol.~52, no.~1, pp. 6--18, 2006.

\bibitem{Tropp:2006vb}
J.~A. Tropp, ``{Just relax: convex programming methods for identifying sparse
  signals in noise},'' \emph{IEEE Transactions on Information Theory}, vol.~52,
  no.~3, pp. 1030--1051, 2006.

\bibitem{tarantola2005inverse}
A.~Tarantola, \emph{Inverse Problem Theory and Methods for Model Parameter
  Estimation}.\hskip 1em plus 0.5em minus 0.4em\relax Society for Industrial
  and Applied Mathematics, 2005.

\bibitem{oakland2003statistical}
J.~Oakland, \emph{Statistical Process Control}, ser. Quality management /
  Butterworth Heinemann.\hskip 1em plus 0.5em minus 0.4em\relax Taylor \&
  Francis, 2003.

\bibitem{BergFriedlander:2008}
E.~van~den Berg and M.~P. Friedlander, ``Probing the pareto frontier for basis
  pursuit solutions,'' \emph{SIAM Journal on Scientific Computing}, vol.~31,
  no.~2, pp. 890--912, 2008.

\end{thebibliography}

\end{document}